\documentclass[11pt,a4paper]{article}
\pdfoutput=1

\usepackage{enumerate}
\usepackage{jheppub,undertilde}
\usepackage{amsthm, amsfonts, amsmath, amssymb}
\usepackage{color, multicol}
\usepackage{graphicx, epsfig}
\usepackage{siunitx}
\usepackage{diagbox}
\usepackage{textcomp,float,gensymb,wrapfig, enumitem,comment,dsfont,subfigure,framed,slashed,appendix}
\usepackage{hyperref}
\usepackage{algorithm,algcompatible}
\usepackage{multirow}
\usepackage{soul}

\definecolor{darkred}{rgb}{0.8,0.2,0.3}
\definecolor{darkblue}{rgb}{0.15, 0.2, .85}

\newcommand{\be}{\begin{equation}}
\newcommand{\ee}{\end{equation}}
\newcommand{\bea}{\begin{eqnarray}}  
\newcommand{\eea}{\end{eqnarray}}  
\newcommand{\met}{E_T^{\rm miss}}
\newcommand{\Bsample}{\mathcal{B}} 
\newcommand{\Tsample}{\mathcal{T}} 
\newcommand{\ts}{\textrm{TS}} 

\hypersetup{colorlinks, citecolor=darkblue, linkcolor=darkblue, urlcolor=darkblue}

\makeatletter
\def\@fpheader{\relax}
\makeatother

\preprint{SISSA 27/2018/FISI}

\title{Guiding New Physics Searches with Unsupervised Learning}

\author{Andrea De Simone,}
\author{Thomas Jacques}

\affiliation{SISSA, via Bonomea 265, 34136 Trieste, Italy}
\affiliation{INFN Sezione di Trieste, via Bonomea 265, 34136 Trieste, Italy}

\emailAdd{andrea.desimone@sissa.it}
\emailAdd{thomas.jacques@sissa.it}

\abstract{
We propose a new scientific application of  unsupervised learning techniques to boost our ability to 
search for new phenomena in data, by detecting discrepancies between two datasets.
These could be, for example, a simulated standard-model background, and an observed dataset containing a potential hidden signal of New Physics. 
We build a statistical test upon a test statistic which measures deviations between two samples, 
using a Nearest Neighbors approach to estimate the local ratio of the density of points. The test is model-independent and non-parametric, requiring no knowledge of the shape of the underlying distributions, and 
it does not bin the data, thus 
retaining full information from the multidimensional feature space. 
As a proof-of-concept, we apply our method to synthetic 
Gaussian data, and to a simulated dark matter signal at the Large Hadron Collider.
Even in the case where the background can not be simulated accurately enough to claim discovery, the technique is a powerful tool to identify regions of interest for further study.
}


\begin{document} 
\maketitle
\flushbottom

\section{Introduction}
\label{sec:intro}

The problem of comparing two independent data samples and looking for deviations is ubiquitous in statistical analyses. It is of particular interest in physics, when addressing the problem of searching for new phenomena in data,  to compare observations with expectations to find discrepancies. In general, one would like to assess (in a statistically sound way) whether the observed experimental data are compatible with the expectations, or there are signals of the presence of new phenomena.

In high-energy physics, although the Standard Model (SM) of particle physics has proved to be extremely successful in predicting a huge variety of elementary particle processes with spectacular accuracy, it is widely accepted that it needs to be extended to account for unexplained phenomena, such as the dark matter of the Universe, the neutrino masses, and more.
The search for New Physics (NP) beyond the SM is the primary goal of the Large Hadron Collider (LHC). 
The majority of NP searches at the LHC are performed to discover or constrain specific models, i.e. specific particle physics extensions of the SM.
Relatively less effort has been devoted to design and carry out strategies for model-independent searches for NP \cite{Aaltonen:2007dg, 
Aaltonen:2008vt, 
CMS-PAS-EXO-08-005, 
CMS-PAS-EXO-10-021, 
Choudalakis:2011qn, 
ATLAS-CONF-2017-001, 
Asadi:2017qon, 
DAgnolo:2018cun, 
Aaboud:2018ufy}.
At the current stage of no evidence for NP in the LHC data, it is of paramount importance to increase the chances of observing the presence of NP in the data. It may even be already there, but it may have been missed by model-specific searches.

Recently, there has been growing interest in applying Machine Learning (ML) techniques to high-energy physics problems, especially using supervised learning
(see e.g. Refs. \cite{Kuusela:2011aa, 
Cranmer:2015bka,
Baldi:2016fzo,
Hernandez2016,
Caron:2016hib,
Bertone:2016mdy,
Weisser:2016cnc,
Dery:2017fap,
Louppe:2017ipp,
Cohen:2017exh, 
Metodiev:2017vrx,
Chang:2017kvc, 
Paganini:2017dwg,
Komiske:2018oaa,
Fraser:2018ieu,
Brehmer:2018eca,
Brehmer:2018kdj, 
Brehmer:2018hga}
and in particular the recent work of Ref.~\cite{DAgnolo:2018cun}  
with which we share some ideas,
although with a very different implementation).
On the other hand, applications of unsupervised learning have been relatively unexplored \cite{Kuusela:2011aa, Andreassen:2018apy, Collins:2018epr}.
In unsupervised learning the data are not labeled, so the presence and the characteristics of new phenomena in the data are not known \emph{a priori}.
One disadvantage of unsupervised learning is that one cannot easily associate a performance metric to the algorithm. Nevertheless, unsupervised methods such as anomaly (or outlier) detection techniques, or clustering algorithms, provide powerful tools to inspect the global and local structures of high-dimensional datasets and discover `never-seen-before' processes.

In this paper, we propose a new scientific application of  unsupervised learning techniques to boost our ability to 
search for new phenomena in data, by measuring the degree of compatibility between two data samples 
(e.g.~observations and predictions).
In particular, we build a statistical test upon a test statistic which measures deviations between two datasets, 
relying on a Nearest-Neighbors technique to estimate the ratio of the local densities of points in the samples.

Generally speaking, there are three main difficulties one may face when trying to carry out a search for the presence of new processes in data: 
(1) a model for the physics describing the new process needs to be assumed, 
which limits the generality of the method;
(2) it is impossible or computationally very expensive to evaluate directly the likelihood function, e.g. due to the complexity of the experimental apparatus;
(3) a subset of relevant features needed to be extracted from the data, otherwise the histogram methods may fail due to the sparsity of points in high-dimensional bins.

A typical search for NP at LHC suffers from all such limitations:
a model of NP (which will produce a signal, in the high-energy physics language) is assumed, the likelihood evaluation is highly impractical, and a few physically motivated variables (observables or functions of observables) are selected to maximize the presence of the signal with respect to the scenario without NP (the so-called background).

Our approach overcomes all of these problems at once, by having the following properties:
\begin{enumerate}
\item it is \textit{model-independent}: it aims at assessing whether or not the observed data contain traces of new phenomena (e.g. due to NP), regardless of the specific physical model which may have generated them;
\item it is \textit{non-parametric}:  it does not make any assumptions about the probability distributions from which the data are drawn, so it is likelihood-free;
\item it is \textit{un-binned}: 
it partitions the feature space of data without using fixed rectangular bins; so it allows one to retain and exploit the information from the full high-dimensional feature space, when single or few variables cannot. 
\end{enumerate}
The method we propose in this paper  is particularly useful when dealing with 
situations where the distribution of data in feature space is almost indistinguishable
from the distribution of the reference (background) model.

Although our main focus will be on high-energy particle physics searches at the LHC, our method 
can be successfully applied in many other situations where one needs to detect incompatibilities between data samples.

The remainder of the paper is organized as follows.
In Section \ref{sec:2ST} we describe the details of the construction of our method and its properties.
In Section \ref{sec:applications} we apply it to case studies with simulated data, both for synthetic Gaussian samples and for a more physics-motivated example related to LHC searches.
We outline some directions for further improvements and extensions of our approach, in Section \ref{sec:extensions}.
Finally, we conclude in Section \ref{sec:conclusion}.

\section{Statistical test of dataset compatibility}
\label{sec:2ST}

In general terms, we approach the problem of measuring the compatibility between datasets sampled from unknown probability densities, 
by first estimating the probability densities and then applying
a notion of functional distance between them. The first task is worked out by performing density ratio estimation using Nearest Neighbors, while the distance between probability densities is chosen to be the Kullback-Leibler  divergence \cite{KullbackLeibler}.
We now describe our statistical test in more detail.

\subsection{Definition of the problem}
\label{subsec:generalities}

Let us start by defining the problem more formally. 
Let $\{\boldsymbol{x}_i | \boldsymbol{x}_i\in\mathbb{R}^D\}_{i=1}^{N_T}$
and $\{\boldsymbol{x}'_i | \boldsymbol{x}'_i\in\mathbb{R}^D \}_{i=1}^{N_B}$ 
be two independent and identically distributed $D$-dimensional samples
drawn independently from the probability density functions (PDFs) $p_T$ and $p_B$, respectively:
\bea
\Tsample&\equiv&\{\boldsymbol{x}_i\}_{i=1}^{N_T} 
\stackrel{\textrm{iid}}{\sim}p_T\,,\\
\Bsample&\equiv&\{\boldsymbol{x}'_i\}_{i=1}^{N_B}
\stackrel{\textrm{iid}}{\sim}p_B\,.
\eea
We will refer to $\Bsample$ as a `benchmark' (or `control' or `reference') sample 
and to $\Tsample$ as a `trial' (or `test') sample.
The $\Tsample, \Bsample$ samples consist of $N_T, N_B$ points, respectively.
The $\mathbb{R}^D$ space where the sample points $\boldsymbol{x}_i,\boldsymbol{x}'_i$ live will be referred to as `feature' space.

The primary goal is to check whether the two samples are drawn from the same PDF, i.e. whether $p_B=p_T$. 
In other words, we aim at assessing whether (and to what significance level) the two samples are compatible with each other.
More formally, 
we want to perform a statistical test of the null hypothesis $\{H_0:p_T=p_B\}$ versus the alternative hypothesis $\{H_1:p_T\neq p_B\}$.

This problem is well-known in the statistics literature as a \textit{two-sample} (or \textit{homogeneity}) test, and many ways to handle it have been proposed.
We want to construct a statistical hypothesis test of dataset compatibility satisfying the properties 1-3 outlined in the introduction.

First, the $\Bsample, \Tsample$ samples are going to be analyzed without any  particular assumptions about the underlying model that generated them (property 1); our hypothesis test does not try to infer or estimate the parameters of the parent distributions, but
it simply 
outputs to what degree the two samples can be considered compatible.

Second, if one is only interested in a location test, such as determining whether the two samples have the same mean or variance, then a $t$-test is often adopted. However, we assume no knowledge about the original PDFs,  and we want to check the equality or difference of the two PDFs as a whole; therefore, we will follow a non-parametric (distribution-free) approach (property 2).

Third, we want to retain the full multi-dimensional information of the data samples, but high-dimensional histograms may result in sparse bins of poor statistical use. The popular Kolmogorov-Smirnov method only works for one-dimensional data,  
and extensions to multi-dimensional data are usually based on binning (for an alternative method that instead reduces the dimensionality of the data to one, see Ref.~\cite{Weisser:2016cnc}).

Alternative non-parametric tests like the Cram\'er-von Mises-Anderson test or the Mann-Withney test require the possibility of ranking the data points in an ordinal way, which may be ill-defined or ambiguous in high-dimensions. Thus, we will employ a different partition of feature space not based on fixed rectangular bins (property 3), which allows us to perform a non-parametric two-sample test in high dimensions.

So, in order to construct our hypothesis test satisfying properties 1-3, we need to build a new test statistic and construct its distribution, as described in the next sections.

\subsection{Test statistic}
\label{subsec:ts}

Since we are interested in measuring the deviation between the two samples, it is convenient to define the ratio of probability densities
to observe the points in the two samples, in the case $p_B\neq p_T$ (numerator)
relative to the case $p_B=p_T$ (denominator)
\be
\lambda
\equiv\frac{\prod_{\boldsymbol{x}'_j\in \Bsample}p_B(\boldsymbol{x}'_j) 
\prod_{\boldsymbol{x}_j\in \Tsample}p_T(\boldsymbol{x}_j)}
{\prod_{\boldsymbol{x}'_j\in \Bsample}p_B(\boldsymbol{x}'_j)
\prod_{\boldsymbol{x}_j\in \Tsample}p_B(\boldsymbol{x}_j)}
=\prod_{\boldsymbol{x}_j\in \Tsample}\frac{p_T(\boldsymbol{x}_j)}{p_B(\boldsymbol{x}_j)}\,.
\ee
The above quantity may also be  thought of as a likelihood ratio.  
However, as we are carrying out a non-parametric test, we prefer not to use this term to avoid confusion.

Now, since the true PDFs $p_{B,T}$ are not known, 
we follow the approach of finding estimators $\hat p_{B,T}$ for the PDFs and evaluate the ratio $\lambda$ on them
\be
\hat\lambda = 
\prod_{\boldsymbol{x}_j\in \Tsample}\frac{\hat p_T(\boldsymbol{x}_j)}{\hat p_B(\boldsymbol{x}_j)}\,.
\ee
We then define our \textit{test statistic} $\ts$ over the trial sample as

\be
\ts(\Bsample, \Tsample)\equiv \log\hat\lambda^{1/|\Tsample|}
=\frac{1}{N_T}\sum_{j=1}^{N_T}\log\frac{\hat p_T(\boldsymbol{x}_j)}
{\hat  p_B(\boldsymbol{x}_j)}\,,
\label{eq:ts}
\ee
where $|\Tsample|=N_T$ is the size of the trial sample.
This test statistic will take values close to zero when $H_0$ is true, and far from zero 
(positively or negatively) when $H_0$ is false. 

The test statistic defined in Eq.~\eqref{eq:ts} is also equal to the estimated Kullback-Leibler (KL) divergence
$\hat D_{\rm KL}(\hat p_T||\hat p_B)$
between the estimated PDFs of trial and benchmark samples, 
with the expectation value replaced by the empirical average 
(see Appendix \ref{app:KL} and in particular Eq.~\eqref{KLestimated}).
The KL divergence plays a central role in information theory and can be interpreted as the relative entropy of a probability distribution with respect to another one. 
Our choice is also motivated by the fact that the log function in Eq.~(\ref{eq:ts}) makes the test
statistic linearly sensitive to small differences between the distributions.
Of course, other choices for the test statistic
are possible, based on an estimated divergence between
distributions other than the KL divergence, e.g. the Pearson squared-error divergence. The exploration of other possibilities is beyond the scope of this paper and is left for future work.

Ultimately, we want to conclude whether or not the null hypothesis can be rejected, with a specified significance level $\alpha$ (e.g. $\alpha=0.05$), therefore we need to associate a $p$-value to the null hypothesis, to be compared with $\alpha$.
To this end, we first need to estimate the PDFs $\hat p_{B,T}$ from the samples, then 
compute the test statistics $\ts_{\rm obs}$ observed on the two given samples. Next, 
in order to evaluate the probability associated with the observed value 
$\ts_{\rm obs}$ of the test statistic, we need to reconstruct its probability distribution 
$f(\ts|H_0)$ under the null hypothesis $H_0$, and finally compute a \textit{two-sided} $p$-value of the null hypothesis.

The distribution of the test statistic is expected to be symmetric around its mean (or median), which in general may not be exactly zero as a finite-sample effect. Therefore, the two-sided $p$-value is simply double the one-sided $p$-value.

\begin{figure}[t]
\centering
\includegraphics[width=0.7\linewidth]{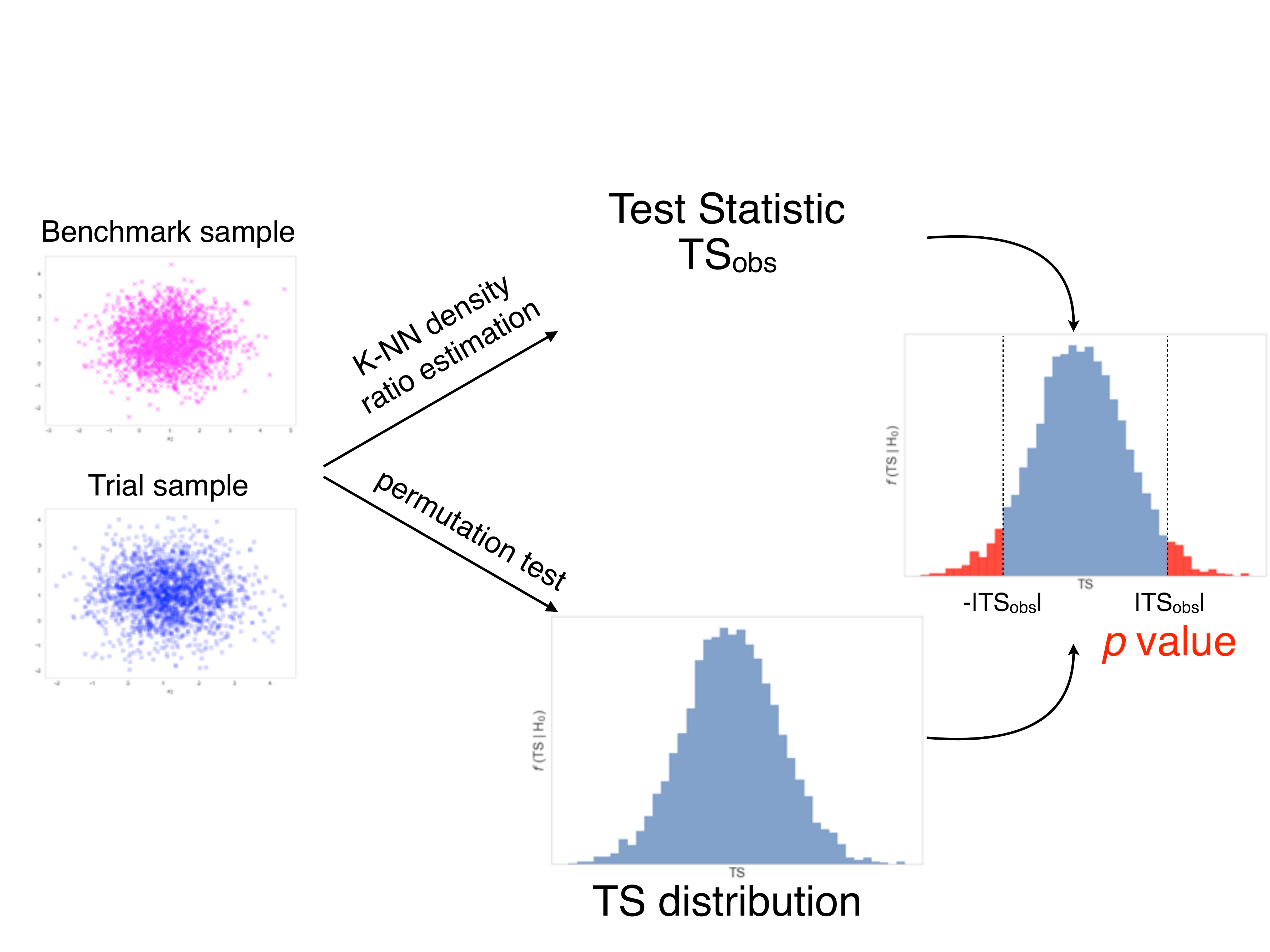}
\caption{
Schematic view of the proposed method to compute the $p$-value of the
null hypothesis that the two samples are drawn from the same probability density.}
\label{fig:schematic_view}
\end{figure}

A schematic summary of the method proposed in this paper is shown in Figure \ref{fig:schematic_view}.
In the remainder of this section we will describe this procedure in detail.

\subsection{Probability density ratio estimator}
\label{subsec:estimator}

We now turn to describing our approach to estimating the
ratio of  probability densities $\hat p_{B}/\hat p_{T}$ needed for the test statistic.
There exist many possible ways to obtain density ratio estimators, e.g. using kernels \cite{SUGIYAMA2011735} 
(see Ref.~\cite{sugiyama_suzuki_kanamori_2012} for a comprehensive review).
We choose to adopt a Nearest-Neighbors (NN) approach \cite{Schilling1986, henze1988, Wang2005,Wang2006,Dasu06, PerezCruz2008, kremer}.

Let us fix an integer $K>0$. For each point $\boldsymbol{x}_j\in \Tsample$, one computes the Euclidean distance\footnote{Other distance metrics may be used, 
e.g. a $L^p$-norm. We do not explore other possibilities here.} $r_{j,T}$ to the $K$th nearest neighbor of $\boldsymbol{x}_j$ in $\Tsample\setminus\{\boldsymbol{x}_j\}$, 
and the Euclidean distance $r_{j,B}$ to the $K$th nearest neighbor of $\boldsymbol{x}_j$ in $\Bsample$. 
Since the probability density is proportional to the density of points, the probability estimates are simply given by the number of points ($K$, by construction) within a sphere of radius $r_{j,B}$ or $r_{j,T}$, divided by the volume of the sphere and the total number of available points.
Therefore, the local nearest-neighbor estimates of the PDFs read
\bea
\hat p_B(\boldsymbol{x}_j) &=& \frac{K}{N_B}\frac{1}{\omega_D r_{j,B}^D}\,,\\
\hat p_T(\boldsymbol{x}_j) &=& \frac{K}{N_T-1}\frac{1}{\omega_D r_{j,T}^D}\,,
\eea
(for any $\boldsymbol{x}_j\in \Tsample$) where
 $\omega_D=\pi^{D/2}/\Gamma(D/2+1)$ is the volume of the unit sphere in 
$\mathbb{R}^D$.
So,  the test statistic defined in Eq.~\eqref{eq:ts} is  simply given by
\be
\ts(\Bsample, \Tsample)=
\frac{D}{N_T}\sum_{j=1}^{N_T}\log\frac{r_{j,B}}{r_{j,T}}
+ \log\frac{N_B}{N_T-1}\,.
\label{eq:ts2}
\ee
The value of the test statistic on the benchmark and trial samples will also be referred to as the `observed' test statistic $\ts_{\rm obs}$.
The NN density ratio estimator  described above has been proved to be consistent and asymptotically unbiased \cite{Wang2005,Wang2006,PerezCruz2008},
i.e.  the test statistic $\ts$ \eqref{eq:ts2} built from the estimated
probability densities converges almost surely to the KL divergence between the true probability densities in the large sample limit $N_{B},N_T\to\infty$.

Two advantages of the NN density ratio estimator are that it easily handles high-dimensional data, and its calculation is relatively fast, especially if $k$-$d$ trees are employed to find the nearest neighbors. As a disadvantage, for finite sample sizes, the estimator \eqref{eq:ts2} retains a small bias, although several methods
has been proposed to reduce it (see e.g. Refs.~\cite{Wang2005, Noh2014BiasRA}). 
Such a residual bias is only related to the 
asymptotic convergence properties of the test statistic to the
estimated KL divergence 
$\hat D_{\rm KL}(\hat p_T||\hat p_B)$, and does not affect the outcome 
and the power of our test in any way.

The use of NN is also convenient as it allows the partition of the feature space not into rectangular bins, but into hyper-spheres of varying radii, making sure they are all populated by data points.

The test statistic $\ts$ in Eq.~\eqref{eq:ts2}, being an estimator of the KL divergence between the two underlying (unknown) PDFs, provides a measure of  dataset compatibility. 
In the construction of $\ts$ we have chosen a particular $K$ as the number of nearest neighbors.  Of course, there is not an \textit{a priori} optimal value of $K$ to choose. In the following analyses we will use a particular choice of $K$, and
we will comment on the possibility of extending the algorithm with adaptive $K$ in 
Section \ref{subsec:adaptive}.

Now that we have a test statistic which correctly encodes the degree of compatibility between two data samples, and its asymptotic properties are ensured by theorems,
we need to associate a probability with the value of the $\ts$ calculated on the given
samples, as described in the next section.

\subsection{Distribution of the test statistic and $p$-value}
\label{subsec:permutation}

In order to perform a hypothesis test, we need to know the distribution of the test statistic $ f(\ts|H_0)$ under the null hypothesis $H_0$, to be used to compute the $p$-value.
Classical statistical tests have well-known distributions of the test statistics, e.g. normal, $\chi^2$ or Student-$t$. In our case, the distribution of $\ts$ is not theoretically known, for finite sample sizes. Therefore, it needs to be estimated from the data samples themselves. We employ the resampling method known as the permutation test \cite{Edgington, vanderVaart} to construct the distribution $f(\ts|H_0)$ of the $\ts$ under the null hypothesis.
It is a non-parametric (distribution-free) method based on the idea of sampling different relabellings of the data, under the assumption  they are coming from the same parent PDF (null hypothesis).

In more detail, the permutation test is performed by first constructing a pool sample by merging the two samples: $\mathcal{U}=\Bsample \cup \Tsample$, then randomly shuffle (sampling without replacement) the elements of $\mathcal{U}$ and assign the first $N_B$ elements to $\tilde \Bsample$, and the remaining $N_T$ elements to $\tilde \Tsample$. Next, one computes the value of the test statistic on $\tilde \Tsample$. 
If one repeats this procedure for every possible permutation (relabelling) of the sample points, one collects a large set of test statistic values under the null hypothesis which provides an accurate estimation of its distribution (exact permutation test). However, it is often impractical to work out all possible permutations, so one typically resorts to perform a smaller number $N_{\rm perm}$ of permutations, which is known as an
approximate (or Monte-Carlo) permutation test. The $\ts$ distribution is then reconstructed from the $N_{\rm perm}$ values of the test statistic obtained by the procedure outlined above.

The distribution of the test statistic under a permutation test is asymptotically normal with zero mean in the large sample limit $N_B,N_T\to \infty$ \cite{vanderVaart}, as a consequence of the Central Limit Theorem.
Furthermore, when the number $N_{\rm perm}$ is large, the distribution of the $p$-value estimator approximately follows a normal distribution with mean $p$ and variance $p(1-p)/N_{\rm perm}$ \cite{EfronTibshirani, Edgington}.
For example, if we want to know the $p$-value in the neighborhood of the significance level $\alpha$ to better than $\alpha/3$, we need  $N_{\rm perm}>9(1-\alpha)/\alpha$, 
which is of the order of 1000 for $\alpha=0.01$.

Once the distribution of the test statistic is reconstructed, it is possible to define the critical region for rejecting the null hypothesis at a given significance $\alpha$,  defined by large enough values of $\ts_{\rm obs}$ such that the corresponding $p$-value is smaller than $\alpha$. 

As anticipated in Section \ref{subsec:ts}, for finite samples the test statistic distribution is still approximately symmetric around the mean, but the latter may deviate from zero.
In order to account for this general case, and give some intuitive meaning to the size of the test statistic, it is convenient to standardize (or `studentize') the $\ts$ to have zero mean and unit variance. Let $\hat\mu, \hat\sigma$ be the mean and the variance of test statistic under the distribution $f(\ts|H_0)$. 
We then transform the test statistic as
\be
\ts \rightarrow \ts' \equiv \frac{\ts - \hat\mu}{\hat \sigma}\,, 
\ee
which is distributed according to 
\be
f'(\ts'|H_0)=\hat\sigma f(\hat\mu+\hat\sigma\ts'|H_0)\,,
\ee
with zero mean and unit variance. With this redefinition, the two-sided $p$-value can be easily computed as  
\be
p = 
2 \int_{|\ts'_{\rm obs}|}^{+\infty} f'(\ts'|H_0)d\ts'\,.
\label{pvalue}
\ee

\subsection{Summary of the algorithm}
\label{subsec:summary}

The pseudo-code of the algorithm for the statistical test 
presented in this paper
is summarized in Table \ref{algo:algo1}.
We implemented it in Python and an open-source package is available on GitHub
\footnote{
\href{https://github.com/de-simone/NN2ST}
{https://github.com/de-simone/NN2ST}}.

\begin{table}[t]
\begin{algorithm}[H]
    \caption{Nearest-Neighbors Two-Sample Test}
  \begin{algorithmic}[1]
    \REQUIRE{Benchmark sample: $\Bsample=\{\boldsymbol{x}'_i|
			\boldsymbol{x}'_i \in \mathbb{R}^{D}\}_{i=1}^{N_B}$,
    		Trial sample: $\Tsample=\{\boldsymbol{x}_j|
            \boldsymbol{x}_j \in \mathbb{R}^{D}\}_{j=1}^{N_T}$}
    \item[\textbf{Input:}] $K, N_{\rm perm}\in \mathbb{N}\setminus\{0\}$.
    \item[\textbf{Output:}] $p$-value of the null hypothesis.
    \item[]
    
    \FOR{$j=1$ to $N_T$}
    	\STATE{$r_{j,B}\leftarrow$ distance of $K$th-NN in $\Bsample$ 
    			from $\boldsymbol{x}_j\in \Tsample$}
    	\STATE{$r_{j,T}\leftarrow$ distance of $K$th-NN in $\Tsample$ 
    			from $\boldsymbol{x}_j\in \Tsample$}
    \ENDFOR
    
    \STATE{$\ts_{\rm obs} \leftarrow 
    	\frac{D}{N_T}\sum_{j=1}^{N_T}
    	\log\frac{r_{j,B}}{r_{j,T}} + 
    	\log{\frac{N_B}{N_T-1}}$}
    	\COMMENT{observed value of test statistic}
	\item[]
        
	\FOR{$n=1$ to $N_{\rm perm}$}
    	\COMMENT{permutation test}
    	\STATE{$\mathcal{U}_n\leftarrow $ randomly reshuffle $\Bsample \cup \Tsample$}
    	\STATE{$\tilde \Bsample\leftarrow$ first $N_B$ elements of $\mathcal{U}_n$ }    
        \STATE{$\tilde \Tsample\leftarrow$ remaining $N_T$ elements of $\mathcal{U}_n$ }    
        \FOR{$j=1$ to $N_T$}
    	\STATE{$\tilde r_{j,B}\leftarrow$ distance of 
        $K$th-NN in $\tilde \Bsample$ 
    			from $\boldsymbol{\tilde x}_j\in \tilde \Tsample$}
    	\STATE{$\tilde r_{j,T}\leftarrow$ distance of 
        $K$th-NN in $\tilde \Tsample$ 
    			from $\boldsymbol{\tilde x}_j\in \tilde \Tsample$}
    	\ENDFOR
    
    	\STATE $ \ts_n \leftarrow 
    		\frac{D}{N_T}\sum_{j=1}^{N_T}
    		\log\frac{\tilde r_{j,B}}{\tilde r_{j,T}}
            + \log{\frac{N_B}{N_T-1}}$
    		\COMMENT{test statistic on permutation $n$}
    \ENDFOR
    \item[]
    
    \STATE $f(\ts|H_0)\leftarrow \{\ts_n\}$ 
            \COMMENT{probability distribution of $\ts$
            under $H_0$}
    \STATE{$\hat\mu,\hat\sigma^2\leftarrow$ mean and variance of $\ts$ 
           under $f$}
    \STATE{$\ts' \leftarrow (\ts - \hat\mu)/\hat \sigma$}
    \STATE{$f'(\ts'|H_0)\leftarrow\hat\sigma f(\hat\mu+\hat\sigma\ts'|H_0)$}
    		\COMMENT{probability distribution of $\ts'$ under $H_0$}
    \STATE{$p\leftarrow 2 \int_{|\ts'_{\rm obs}|}^{+\infty} f'(\ts'|H_0)d\ts'$}

  \end{algorithmic}
\end{algorithm}
\caption{Pseudo-code for the two-sample test algorithm, using nearest neighbors density ratio estimation.}
\label{algo:algo1}
\end{table}

\subsection{Extending the test to include uncertainties}
\label{subsec:ext_uncert}

So far we have assumed that both $\Bsample$
and $\Tsample$ samples are precisely known.
However, in several situations of physical interest
this may not be the case, as the features may be known only with some uncertainty,
e.g. when the sample points come from physical measurements.
There can be several factors affecting the precision with which each sample point is known, for instance
systematic uncertainties (e.g. the smearing effects of the detector response) and the limited accuracy of the background (Monte-Carlo simulation), which may be particularly poor in some regions of the feature space.

Of course, once such uncertainties are properly taken into account, we expect a degradation of the results of the statistical test described in the previous sections, leading to weaker conclusions about the rejection of the null hypothesis.

Here we describe a simple and straightforward 
extension of the method described in this section, to account
for uncertainties in the positions of the sample points.
We consider the test
statistic itself as a random variable, which is a sum of the test statistic $\ts$ defined in Section \ref{subsec:ts}, and
computed on the original $\Bsample,\Tsample$ samples, and an uncertainty fluctuation (noise) $U$,
originating when each point of $\Bsample$ (or $\Tsample$ or both) is shifted by a random vector: $\ts_u=\ts+U$.
The trial and benchmark samples with uncertainties are then given by 
\bea
\Tsample_u&=&
\{\boldsymbol{x}_i+\Delta\mathbf{x}_i\}_{i=1}^{N_T}\,, \\
\Bsample_u&=&
\{\boldsymbol{x}'_i+\Delta\mathbf{x}'_i\}_{i=1}^{N_B}\,,
\eea
which represent a point-wise random shift, where
the error samples $\Delta\mathbf{x}_i,\Delta\mathbf{x}_i'\in\mathbf{R}^D$ 
are independent random variables drawn from the same distribution, according to the expected (or presumed) distribution of uncertainties in the features, e.g. zero-mean multivariate Gaussians.

Next, one can compute the test statistic on the `shifted'
samples as
\be
\ts_u\equiv\ts(\Bsample_u, \Tsample_u)=
\ts(\Bsample, \Tsample) + U\,.
\ee
Since the $\ts$ computed on the original $\Bsample, \Tsample$ samples
is given by the observed value $\ts_{\rm obs}$, 
the value of $U$ for any random samplings of the error samples
is simply $U=\ts(\Bsample_u, \Tsample_u)-\ts_{\rm obs}$.
By repeating the calculation of $U$ many ($N_{\rm iter}$) times, 
each time adding a random noise to $\Bsample$ (or $\Tsample$ or both) we can reconstruct its probability distribution $f(U)$, 
which is asymptotically normal with zero mean in the large-sample limit
$N_B,N_T\to\infty$.

The resulting distribution of the test statistic $\ts_u$, 
being the sum of two i.i.d. random variables, is then given by the convolution of the distribution $f(\ts|H_0)$, 
computed via permutation test on $\Bsample,\Tsample$, 
and the distribution $f(U)$ with mean set to zero. 
This is motivated by the desire to eliminate the  bias in the mean of the 
distribution of $U$ coming from finite-sample effects.
As a result of this procedure, the distribution $f(\ts_u|H_0)$
will have the same mean as $f(\ts|H_0)$ but a larger variance.

The $p$-value of the test is computed from $\ts_{\rm obs}$ with the same steps as described in Section \ref{subsec:permutation}, but
with the distribution of the test statistic with uncertainties
given by $f(\ts_u|H_0)$,
rather than $f(\ts|H_0)$.
Since $f(\ts_u)$ has larger variance than $f(\ts)$, 
the $p$-value will turn out to
be larger, therefore the equivalent significance $Z$ will be smaller. 
This conclusion agrees with the expectation that the inclusion of uncertainties leads to a degradation of the power of the test.

The summary of the algorithm to compute the distribution $f(U)$
can be found in  Table \ref{algo:algo2}.
Once $f(U)$ is computed, it needs to be convolved with
$f(\ts|H_0)$, which was previously found via permutation test, as described in Section \ref{subsec:permutation}, to provide 
the distribution of the test statistic with uncertainties
needed to compute the $p$-value.

\begin{table}[t]
\begin{algorithm}[H]
    \caption{Distribution of the test statistic noise}
  \begin{algorithmic}[1]
  
    \REQUIRE{Benchmark sample: $\Bsample=\{\boldsymbol{x}'_i|
			\boldsymbol{x}'_i \in \mathbb{R}^{D}\}_{i=1}^{N_B}$,
    		Trial sample: $\Tsample=\{\boldsymbol{x}_j|
            \boldsymbol{x}_j \in \mathbb{R}^{D}\}_{j=1}^{N_T}$}
            
    \item[\textbf{Input:}] $K, N_{\rm iter}\in 
    				\mathbb{N}\setminus\{0\}$
                    
    \item[\textbf{Input:}] 
    			$F_{\Bsample}(\mathbf{x}),
                 F_{\Tsample}(\mathbf{x})$:
                distributions of feature uncertainties
                 for $\Bsample, \Tsample$ samples
                 
    \item[\textbf{Output:}] $f(U)$: distribution 
    		of the test statistic noise $U$
    \item[]
    
	\STATE{$\ts_{\rm obs}\leftarrow\ts(\Bsample, \Tsample)$}
    \COMMENT{observed value of test statistic}
            
    \FOR{$j=1$ to $N_{\rm iter}$}
    
        \STATE{$\mathcal{E}_\Tsample=
    		\{\Delta \mathbf{x}_i\}_{i=1}^{N_T}$ 
            randomly drawn from $F_{\Tsample}(\mathbf{x})$}
            
	    \STATE{$\mathcal{E}_\Bsample=
    		\{\Delta \mathbf{x}_i'\}_{i=1}^{N_B}$ 
            randomly drawn from $F_{\Bsample}(\mathbf{x})$}            
	
    	\STATE{$\Tsample_u\leftarrow
        	\Tsample+\mathcal{E}_\Tsample$ }
            \COMMENT{point-wise sum}
        \STATE{$\Bsample_u\leftarrow
        	\Bsample+\mathcal{E}_\Bsample$ }
            \COMMENT{point-wise sum}
        
    	\STATE{$\ts_{u}\leftarrow\ts(\Bsample_u, \Tsample_u)$}
        \STATE{$U_j\leftarrow\ts_u-\ts_{\rm obs}$}    
    \ENDFOR
        
    \STATE $f(U)\leftarrow \{U_j\}$ 
            \COMMENT{distribution of $U$}
    
  \end{algorithmic}
\end{algorithm}
\caption{Pseudo-code for the algorithm to find the distribution
$f(U)$ of the test statistic noise $U$.
}
\label{algo:algo2}
\end{table}

\section{Applications to simulated data}
\label{sec:applications}

\subsection{Case study: Gaussian samples}
\label{subsec:2dgaussians}

\begin{figure}
\centering
\includegraphics[width=0.7\linewidth]{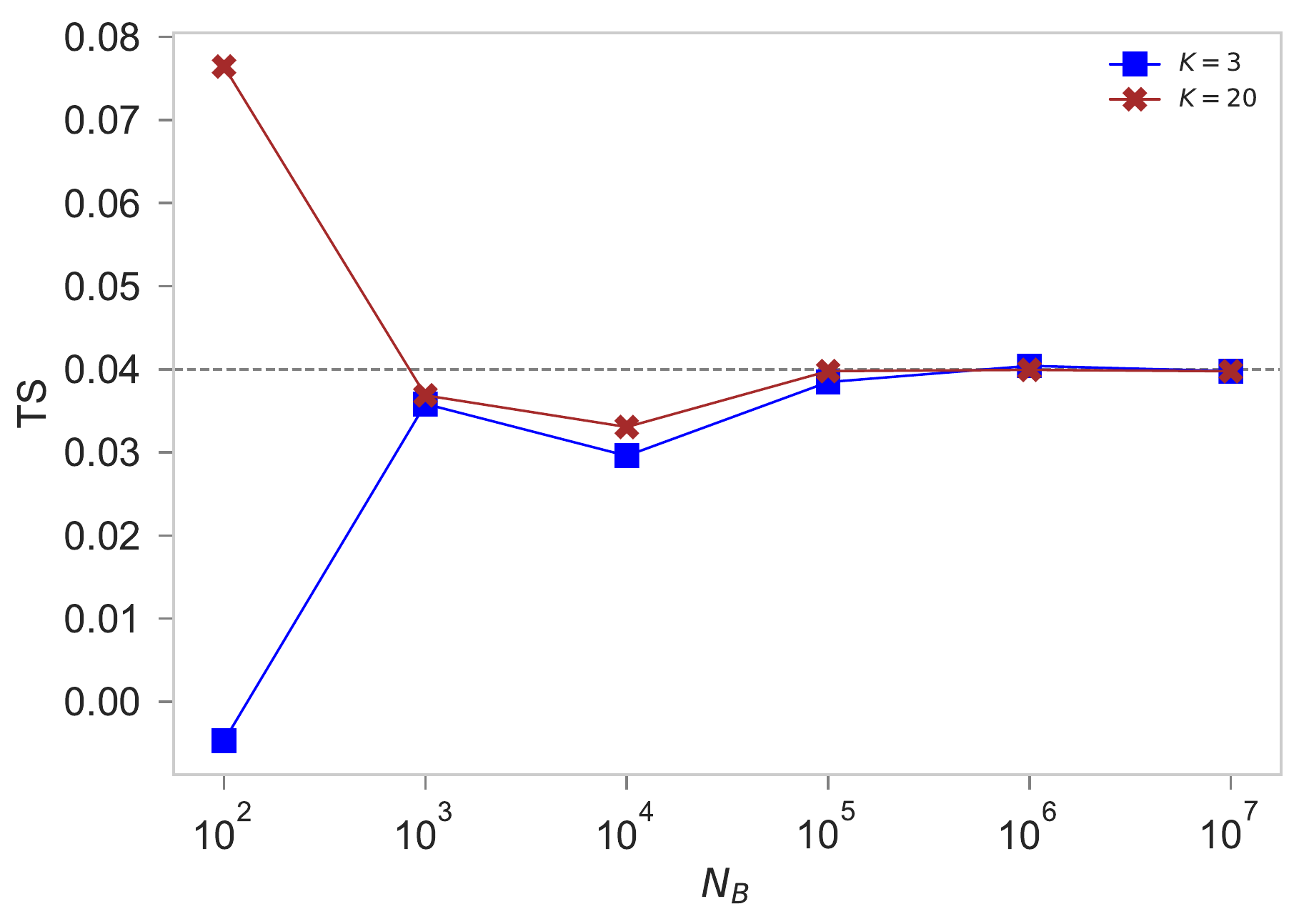}
\caption{
Convergence of the test statistic to the exact KL divergence (dashed horizontal line) between two 2-dimensional Gaussian distributions, in the large-sample limit.
The $\Bsample$,$\Tsample$ samples have the same size $N_B=N_T$, 
and they are sampled from 2-dimensional Gaussian distributions with
$\boldsymbol{\mu}_B=1.0_2$, 
$\boldsymbol{\mu}_T=1.2_2$,
$\Sigma_B=\Sigma_T=\boldsymbol{I}_2$.
Two different choices for the number of nearest neighbors are shown: $K=3$ (blue squares) and $K=20$ (red crosses).
}
\label{convergence_2Dgaussians}
\end{figure}

As a first case study of our method let us suppose
we know the original distributions from which
the benchmark and trial samples are randomly drawn.
For instance, let us consider the 
multivariate Gaussian distributions of dimension $D$ defined by  mean vectors $\boldsymbol{\mu}_{B,T}$
and covariance matrices $\Sigma_{B,T}$:
\be
p_B=\mathcal{N}(\boldsymbol{\mu}_B, \Sigma_B)\,,
\qquad
p_T=\mathcal{N}(\boldsymbol{\mu}_T, \Sigma_T)\,.
\ee
In this case, the KL divergence can be computed analytically (see Eq.~\eqref{KL_gaussians}).
In the large sample limit, 
we recover that the test statistic converges
to the true KL divergence between the PDFs (see Figure
\ref{convergence_2Dgaussians} and Appendix \ref{app:KL}).
Of course, the comparison is possible because
we knew the parent PDFs $p_B, p_T$.

For our numerical experiments we fix the benchmark $\Bsample$ sample by the parameters
$\boldsymbol{\mu}_B=1_D$,
$\Sigma_B=\boldsymbol{I}_D$,
and we construct 4 different trial samples 
$\Tsample_{G0}, \Tsample_{G1}, \Tsample_{G2}, \Tsample_{G3}$ drawn by Gaussian distributions
whose parameters are defined in Table \ref{tab:datasets}. 
Each sample consists of 20\,000 points randomly drawn
from the Gaussian distributions defined above.
Notice that the first trial sample 
$\Tsample_{G0}$ is drawn from the same distribution as the benchmark sample.

\begin{table}[t]
\centering
\begin{tabular}{|c|c|c|}
\hline
Dataset &  $\boldsymbol{\mu}$ & 
  $\Sigma$ \\
\hline
$\Bsample$ & $1_D$ & $\boldsymbol{I}_D$  \\
\hline
$\Tsample_{G0}$ & $1_D$ & $\boldsymbol{I}_D$  \\
\hline
$\Tsample_{G1}$ & $1.12_D$ &$\boldsymbol{I}_D$   \\
\hline
$\Tsample_{G2}$ & $1_D$ & 
$\left(
\begin{array}{c|c}
\begin{matrix}
0.95 & 0.1\\
0.1 & 0.8
\end{matrix} & \boldsymbol{0}\\
\hline
\boldsymbol{0}& \boldsymbol{I}_{D-2}
\end{array}
\right)$\\
\hline
$\Tsample_{G3}$ & $1.15_D$& $\boldsymbol{I}_D$ \\
\hline
\end{tabular}
\caption{
Definition of the
Gaussian datasets used for the numerical experiments.
Each sample consists of $N_B=N_T=20\,000$ points randomly drawn
from $D$-dimensional Gaussian distributions $\mathcal{N}(\boldsymbol{\mu}, \Sigma)$.
}
\label{tab:datasets}
\end{table}

\begin{table}[t]
\centering
\begin{tabular}{|c|c|c|c|c|c|c|c|}
\hline
 \multirow{2}{*}{Trial Dataset} & \multicolumn{2}{c|}{$D=2$} &
   \multicolumn{2}{c|}{$D=5$} &  \multicolumn{2}{c|}{$D=10$} \\ 
  \cline{2-7}
  & $p$-value & $Z$ & $p$-value & $Z$  & $p$-value & $Z$ \\
\hline
$\Tsample_{G0}$ 
& $8.2\cdot 10^{-1}$ & 0.2 $\sigma$
& $6.9\cdot 10^{-1}$ & 0.4 $\sigma$
& $6.9\cdot 10^{-1}$ & 0.4 $\sigma$ \\
\hline
$\Tsample_{G1}$ 
& $2.8\cdot 10^{-2}$ & 2.2 $\sigma$
& $1.5\cdot 10^{-7}$ & 5.2 $\sigma$
& $3.6\cdot 10^{-13}$ & 7.3 $\sigma$\\
\hline
$\Tsample_{G2}$ 
& $4.0\cdot 10^{-4}$ & 3.5 $\sigma$
& $8.8\cdot 10^{-8}$ & 5.3 $\sigma$
& $9.4\cdot 10^{-9}$ & 5.7 $\sigma$ \\
\hline
$\Tsample_{G3}$
& $1.2\cdot 10^{-6}$ & 4.9 $\sigma$ 
& $1.4\cdot 10^{-19}$ & 9.1 $\sigma$ 
& $1.9\cdot 10^{-30}$ & 11.5 $\sigma$ \\
\hline
\end{tabular}
\caption{
Summary of the results comparing $\Bsample$ with 4 trial samples, for different
dimensionality $D$. 
The samples are  defined in Table \ref{tab:datasets}.
We set $K=5$ and $N_{\rm perm} = 1000$.
}
\label{tab:gaussian-results}
\end{table}

As is customary, we associate an equivalent Gaussian significance $Z$ to a given (two-sided) $p$-value as: $Z\equiv\Phi^{-1}(1-p/2)$, where $\Phi$ is the cumulative distribution  of a standard (zero-mean, unit-variance) one-dimensional Gaussian distribution.
In Table \ref{tab:gaussian-results} we show the $p$-values and the corresponding $Z$ significance
of the statistical tests for different dimensions $D$.
The results are interpreted as follows.
For $D=2$, the first two trial samples $\Tsample_{G0}, \Tsample_{G1}$ are not distinguished from the benchmark $\Bsample$ at more than 99\%CL ($p>0.01$), while $\Tsample_{G2}, \Tsample_{G3}$ 
are distinguished ($p\leq 0.01$, or equivalently 
$Z\geq 2.6\sigma$). Therefore, one would reject the null hypothesis  at more than 99\%\,CL and conclude
that the PDFs from which $\Tsample_{G2}, \Tsample_{G3}$ are drawn are different from the benchmark PDF $p_B$.
It is remarkable that our statistical test is able to reject the null hypothesis with a large significance of $4.9\sigma$ for two random samples $\Bsample, \Tsample_{G3}$ drawn from 
2-dimensional distributions which only differ 
by a shift of the mean by 15\% along each dimension.
For higher dimensionality of the data, the
discriminating power of the test increases, 
and the null hypothesis is rejected at more than $5\sigma$ significance for all trial samples 
$\Tsample_{G1}, \Tsample_{G2}, \Tsample_{G3}$.
The running time to compute the $p$-value on a standard laptop for two 2-dimensional samples of 20\,000 points each, and for 1000 permutations, was about 2 minutes. The running time scales linearly with the number of permutations.

\begin{figure}
\centering
\includegraphics[width=0.45\linewidth]{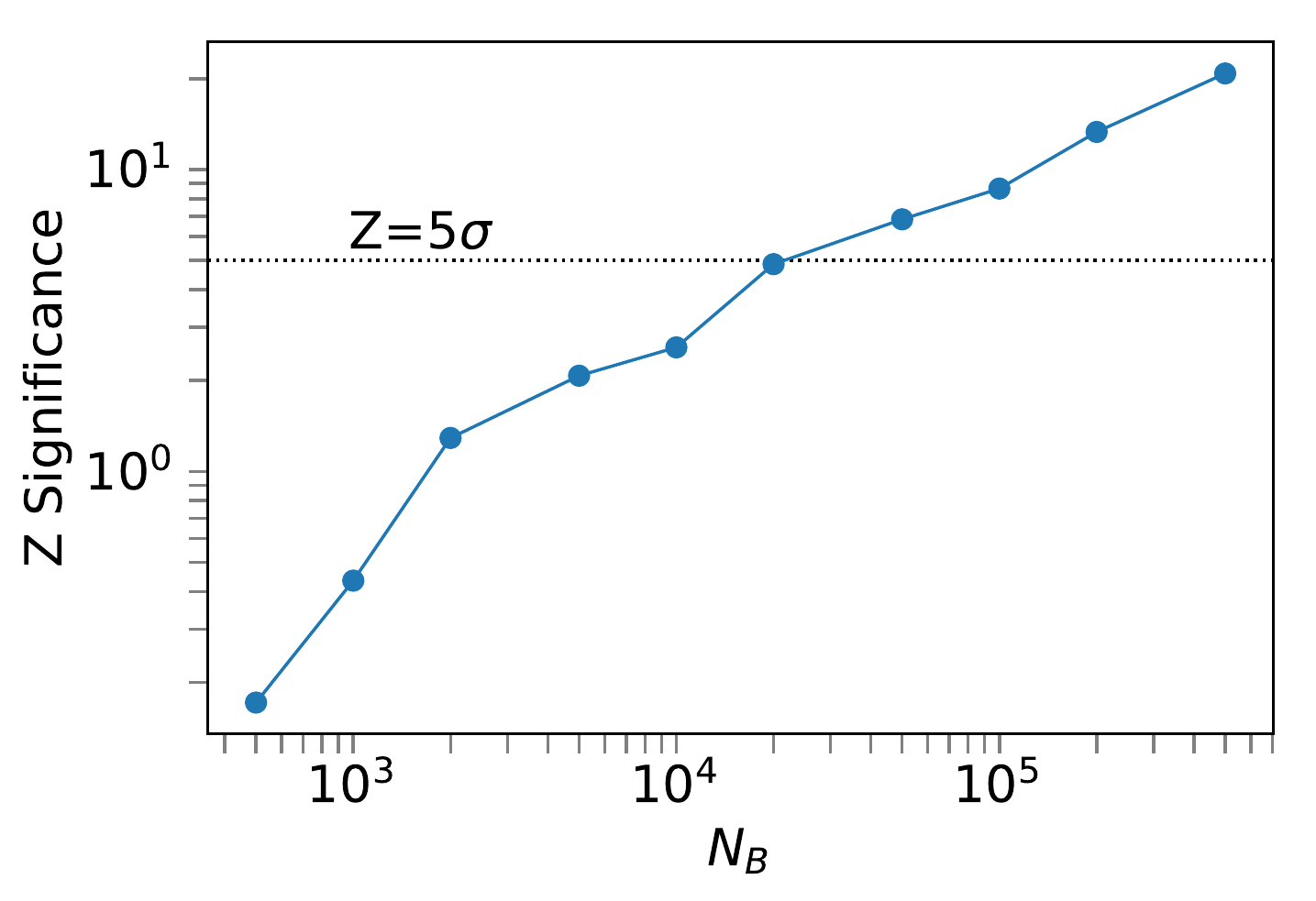}
\includegraphics[width=0.45\linewidth]{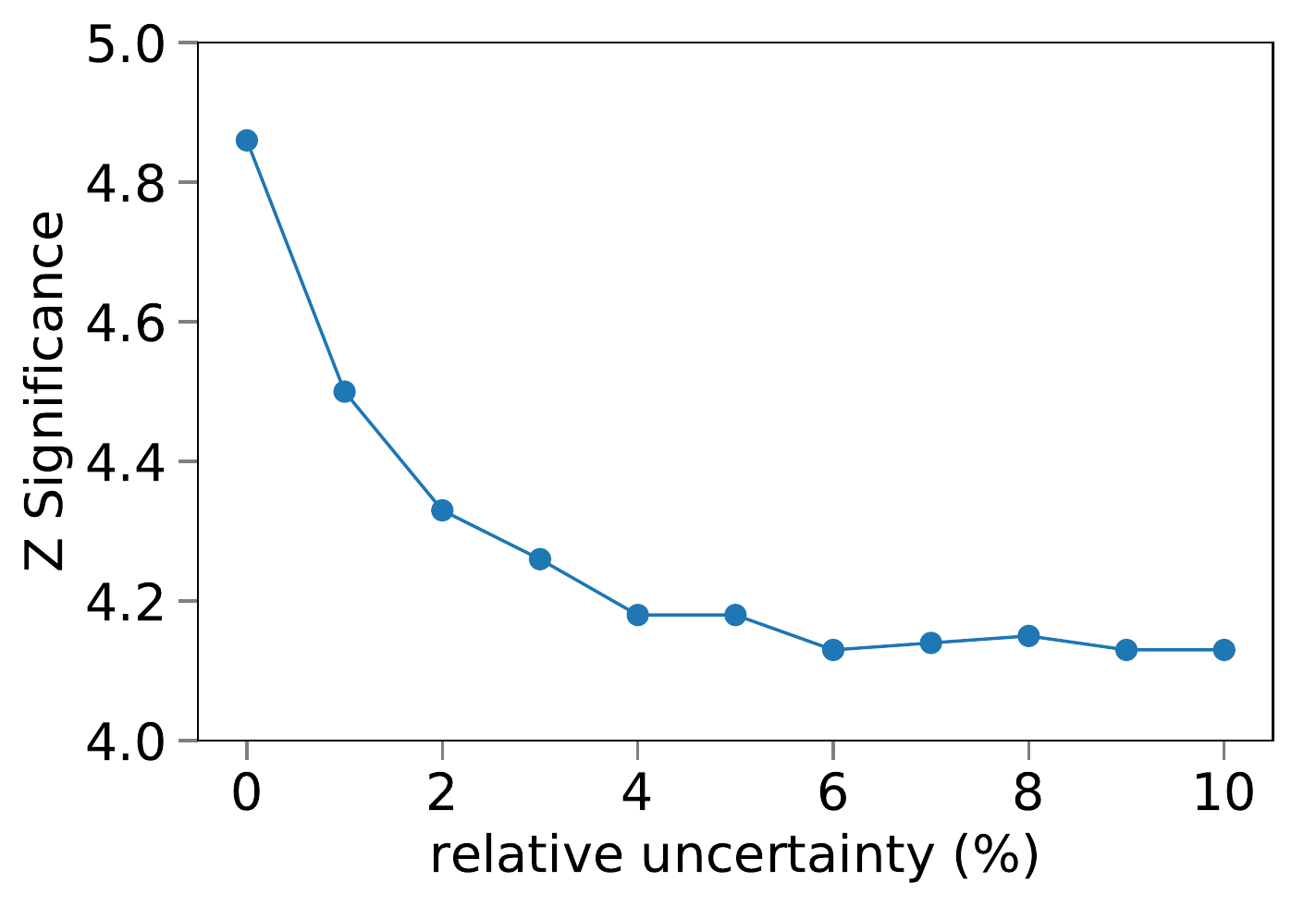}
\caption{
We compare $\Bsample=\Tsample_{G0}$ and $\Tsample=\Tsample_{G3}$, as defined
in
Table \ref{tab:datasets}, with $D=2$, using $K=5$ and $N_{\rm perm}=1000$.
The $\Bsample$,$\Tsample$ samples have the same size $N_B=N_T$.
\emph{Left panel:}
The $Z$ significance of the test for different sample sizes.
\emph{Right panel:}
The $Z$ significance for 
different relative uncertainties added to $\Bsample$ only, 
with fixed $N_B=N_T=20\,000$.
The $U$ distribution has been computed with $N_{\rm iter}=1000$ random samplings from the distribution of feature uncertainties.
}
\label{Z_vs_Nsamples}
\end{figure}

The number of sample points ($N_{B,T}$) plays an important role. 
As an example, we sampled the same datasets $\Bsample, \Tsample_{G0}, \Tsample_{G1}, \Tsample_{G2}, \Tsample_{G3}$ with $N_B=N_T=2000$ points, i.e. ten times less points than for the cases shown in Table \ref{tab:gaussian-results}. The results for the equivalent significance for $\Tsample_{G0}, \Tsample_{G1}, \Tsample_{G2}, \Tsample_{G3}$ 
with $D=2$
are $Z=1.4 \sigma$, $1.9 \sigma$, $1.9 \sigma$, $2.3 \sigma$, respectively.
Clearly, the test is not able to reject the null hypothesis at more than 99\%CL (the $p$-value is never below 0.01,  
or equivalently $Z<2.6\sigma$) in none of the cases.
As another illustration of this point, we run the statistical test for $\Bsample=\Tsample_{G0}$ vs $\Tsample=\Tsample_{G3}$ for $D=2$ and different sample sizes $N_B=N_T$, 
and show the resulting $Z$ significance in Figure \ref{Z_vs_Nsamples} (left panel). 
We find that for  $N_B\leq 10^4$, the test is not able to reject the null hypothesis at more than 99\%CL.
Therefore, the power of our statistical test increases for larger sample sizes, as expected since bigger samples lead to more accurate approximations of the original PDFs.

We have also studied the power performance of our statistical test with respect to parametric competitors.
We ran 200 tests of two samples drawn 
from multivariate Gaussian distributions with $D=1,2,5$, with sample sizes $N_B=N_T=100$, and computed the approximated power as the fraction of runs where the null hypothesis is rejected with
significance level 5\%
($p<0.05$).
We considered normal location alternatives, 
with $\Bsample\sim \mathcal{N}(0_D, \boldsymbol{I}_D)$
and $\Tsample\sim \mathcal{N}(\Delta_D, \boldsymbol{I}_D)$, where $\Delta$ varies
from $0.05$ to $1.0$.
As competitor, we choose the Student's $t$-test (or its generalization Hotelling's $T^2$-test, for $D>1$).
We find that our test shows 
a power comparable to its competitor, 
in some cases lower than that
by at most a factor of 3, which is satisfactory given that the $T^2$-test is parametric
and designed to spot location differences.

Next, we run the statistical test by including uncertainties,
as described in Section \ref{subsec:ext_uncert}.
For the uncertainties, we assume uncorrelated Gaussian noise, 
so the covariance matrix of the uncertainties is a $D$-dimensional 
diagonal matrix $\textrm{diag}(\sigma_1^2,\ldots, \sigma_D^2)$
where each eigenvalue is proportional to the relative uncertainty
$\epsilon$ of the component $x_i$ of the sample point $\mathbf{x}$: 
$\sigma_i=\epsilon x_i$.

In Figure \ref{Z_vs_Nsamples} (right panel) we show how the
significance of rejecting the null hypothesis degrades
once uncorrelated relative uncertainties are added to the 
$\Bsample$ sample. 
For $D=2$, the initial $4.9\,\sigma$ when comparing
$\Bsample=\Tsample_{G0}$ and $\Tsample=\Tsample_{G3}$
without noise goes down to about $4.1\,\sigma$ with 10\%
relative error.

\subsection{Case study: Monojet searches at LHC}
\label{subsec:monojet}

A model-independent search at the LHC for physics Beyond the Standard Model (BSM), such as Dark Matter (DM),  has been elusive
\cite{
CMS-PAS-EXO-08-005, 
CMS-PAS-EXO-10-021, 
Choudalakis:2011qn, 
ATLAS-CONF-2017-001}.
Typically it is necessary to simulate the theoretical signal in a specific model, and compare with data to test whether the model is excluded. The signal-space for DM and BSM physics in general  is enormous, and despite thorough efforts, the possibility exists that a signal has been overlooked. The compatibility test described in Section \ref{sec:2ST} is a promising technique to overcome this challenge, as it can search for deviations between the expected simulated Standard Model signal and the true data, without any knowledge of the nature of the new physics. 

In a real application of our technique by experimental collaborations, the benchmark dataset $\Bsample$  will be a simulation of the SM background, while the trial dataset $\Tsample$ will consist of real measured data, potentially containing an unknown mix of SM and BSM events. As a proof-of-principle, we test whether our method would be sensitive to a DM signature in the monojet channel. For our study, both $\Bsample$ and $\Tsample$ will consist of simulated SM events (`background'), however $\Tsample$ will additionally contain injected DM events (`signal'). The goal is to determine whether the algorithm is sensitive to differences in $\Bsample$ and $\Tsample$ caused by this signal.

\subsubsection*{Model and simulations}
The signal comes from a standard simplified DM model 
(see e.g. Ref.~\cite{DeSimone:2016fbz} for a review) with Fermion DM $\chi$ and an $s$-channel vector $Z'$ mediator \cite{Boveia:2016mrp, Albert:2017onk}. Our benchmark parameters are $g_\chi = 1$, $g_q = 0.1$, $g_\ell = 0.01$, in order to match the simplified model constraints from the ATLAS summary plots \cite{vector_summary}. We use a DM mass of 100 GeV, and mediator masses of (1200, 2000, 3000) GeV, in order to choose points that are not yet excluded but could potentially be in the future \cite{vector_summary}.

Signal and background events are first simulated using MG5\_aMC@NLO v$2.6.1$ \cite{Alwall:2014hca} at center-of-mass energy $\sqrt{s}=13$ TeV, with a minimal cut of $\met > 90$ GeV, to emulate trigger rather than analysis cuts. We use Pythia 8.230 \cite{Sjostrand:2014zea} for hadronization and Delphes 3.4.1 \cite{delphes} for detector simulation. 
The so-called `monojet' signal consists of events with missing energy from DM and at least one high-$p_T$ jet. The resulting signal cross-section is $\sigma_{\rm signal} = (20.4,\,3.8,\,0.6)$ pb for $M_{\rm med} = (1200,\,2000,\,3000)$ GeV respectively.
For the background samples, we simulate 40\,000 events of the leading background, $Z\rightarrow \nu \bar \nu + n j$ where $n$ is 1 or 2, resulting in a cross section of $\sigma_{\rm background} = 202.6$ pb.

The Delphes ROOT file is converted to LHCO and a feature vector is extracted with Python for each event, consisting of $p_T$ and $\eta$ for the two leading jets; the number of jets; missing energy $\met$; Hadronic energy $H_T$; and $\Delta \phi$ between the leading jet and the missing energy. Together this gives an 8-dimensional feature vector $(D=8)$, which is scaled to zero-mean unit-variance based on the mean and variance of the background simulations. This feature vector is chosen to capture sufficient information about each event while keep running time of the algorithm reasonable.  Other choices of the feature vector could be chosen to capture different aspects of the physical processes, including higher- or lower-level features, such as raw particle 4-vectors. Application of high-performance computing resources would allow the feature vector to be enlarged, potentially strengthening results. A full study of the choice of feature vector is left to future work.
Our simulation technique is simple and designed only as a proof of principle; we do not include sub-leading SM backgrounds, nor full detector effects, adopting a generic Delphes profile.

\subsubsection*{Test Statistic distribution under null hypothesis}

Following the technique described in Section~\ref{sec:2ST}, for each of the 3 considered points in signal model parameter space, we first construct an empirical distribution of the test statistic under the null hypothesis, $f(\ts|H_0)$, and we then measure $\ts_{\rm obs}$ and compute the $p$-value to determine the compatibility of the datasets.
We choose $K=5$ and $f(\ts|H_0)$ is constructed over $N_{\rm perm} = 3000$.

\begin{figure}
\centering
\includegraphics[width=0.6\linewidth]{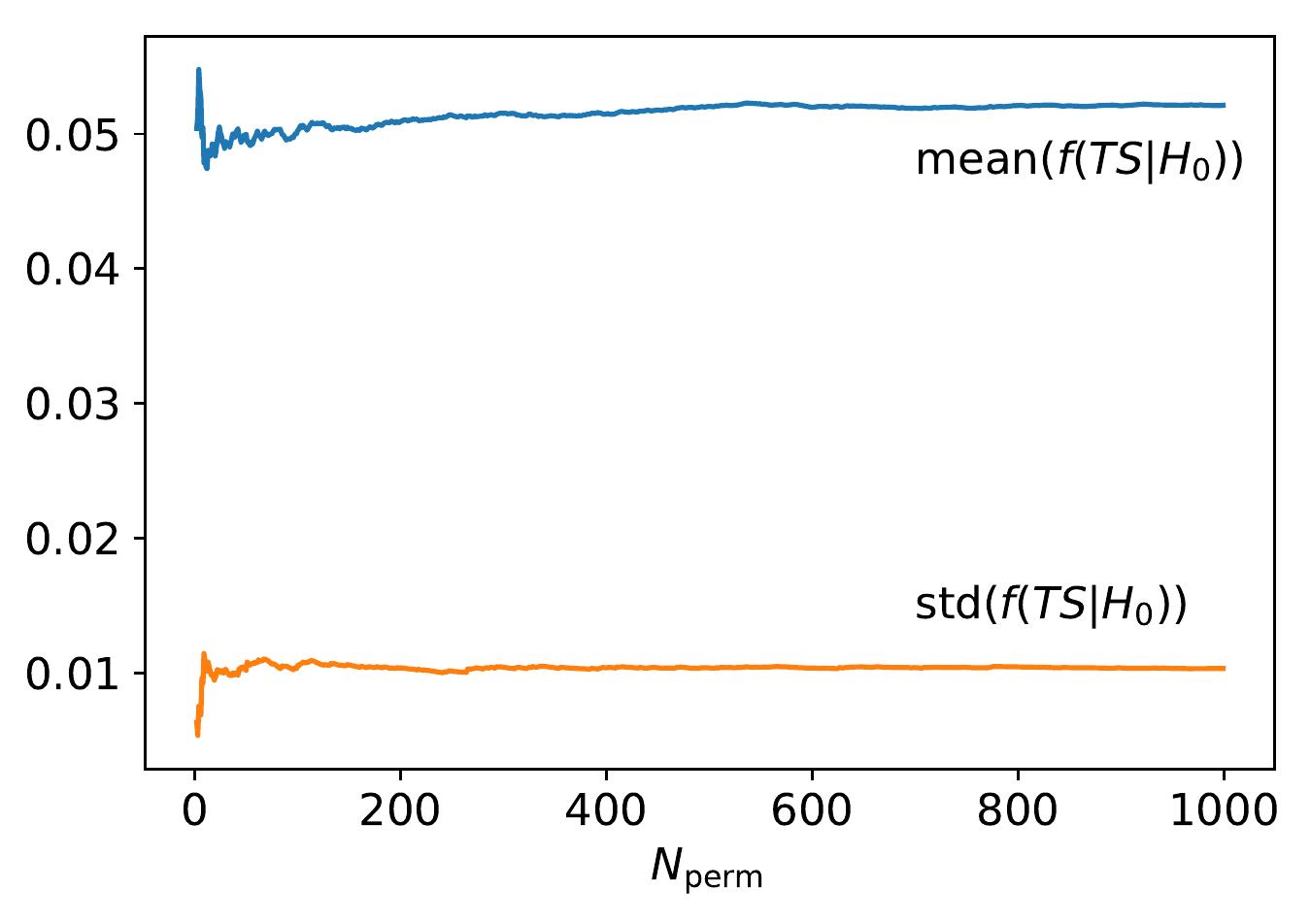}
\caption{
Effect of $N_{\rm perm}$ on the null-hypothesis test statistic for the monojet study with $\Tsample_2$.
}
\label{fig:TSvsNPerm-monojet}
\end{figure}

The pool sample $\Bsample \cup \Tsample$ consists of the 40\,000 background events, along with a number of signal events proportional to the signal cross-section. We define $\Bsample$ and $\Tsample$ as having an equal number of background events, so that $N_{\rm signal} = 20\,000 \times \sigma_{\rm signal}/\sigma_{\rm background}$, $N_T = 20\,000 + N_{\rm signal}$. 
The resulting distribution of TS under the null hypothesis is shown in Fig.~\ref{fig:TS-monojet}. The simulations are relatively fast, taking approximately an hour per 1000 permutations on a standard laptop, although computation time grows as a power-law with the number of events, such that further optimization and high-performance computing resources will be a necessity for application to real LHC data with many thousands of events. The statistics of $f(\ts|H_0)$ converge quickly, as shown in Fig.~\ref{fig:TSvsNPerm-monojet}, consistent with the discussion of $N_{\rm perm}$ in Section \ref{subsec:permutation}, and showing that $N_{\rm perm}$ is more than sufficient.

Note that since $\tilde{\Bsample}, \, \tilde{\Tsample}$ are chosen from permutations of $\Bsample \cup \Tsample$, it is not necessary to specify how the 40\,000 background events are divided between $\Bsample$ and $\Tsample$; It is only necessary to specify $N_B$ and $N_T$ at this point.

\subsubsection*{Observed Test Statistic}
To test whether the null hypothesis would be excluded in the event of an (otherwise unobserved) DM signal hiding in the data, we calculate $\ts_{\rm obs}$ using $\Bsample$ containing only background, and $\Tsample$ containing background plus a number of signal events proportional to the relative cross section. 
In a practical application of this technique by the experimental collaborations, $\Bsample$ would instead correspond to background simulations, while $\Tsample$ would be the real-world observation; therefore only one measurement of $\ts_{\rm obs}$ would be performed. 

However, in our case the distribution of TS under the null hypothesis is insensitive to the way the 40\,000 background events are divided between $\Bsample$ and $\Tsample$.
Therefore we can simulate multiple real-world measurements of $\ts_{\rm obs}$ by dividing the 40\,000 background events between $\Bsample$ and $\Tsample$ in different permutations (always keeping 20\,000 background events in each sample).
This allows us to be more robust: since $\ts_{\rm obs}$ is itself a random variable, multiple measurements of $\ts_{\rm obs}$ allows us to avoid the claim of a small $p$-value, when in reality the algorithm may not be sensitive to a small signal.

\begin{figure}
\centering
\includegraphics[width=0.32\linewidth]{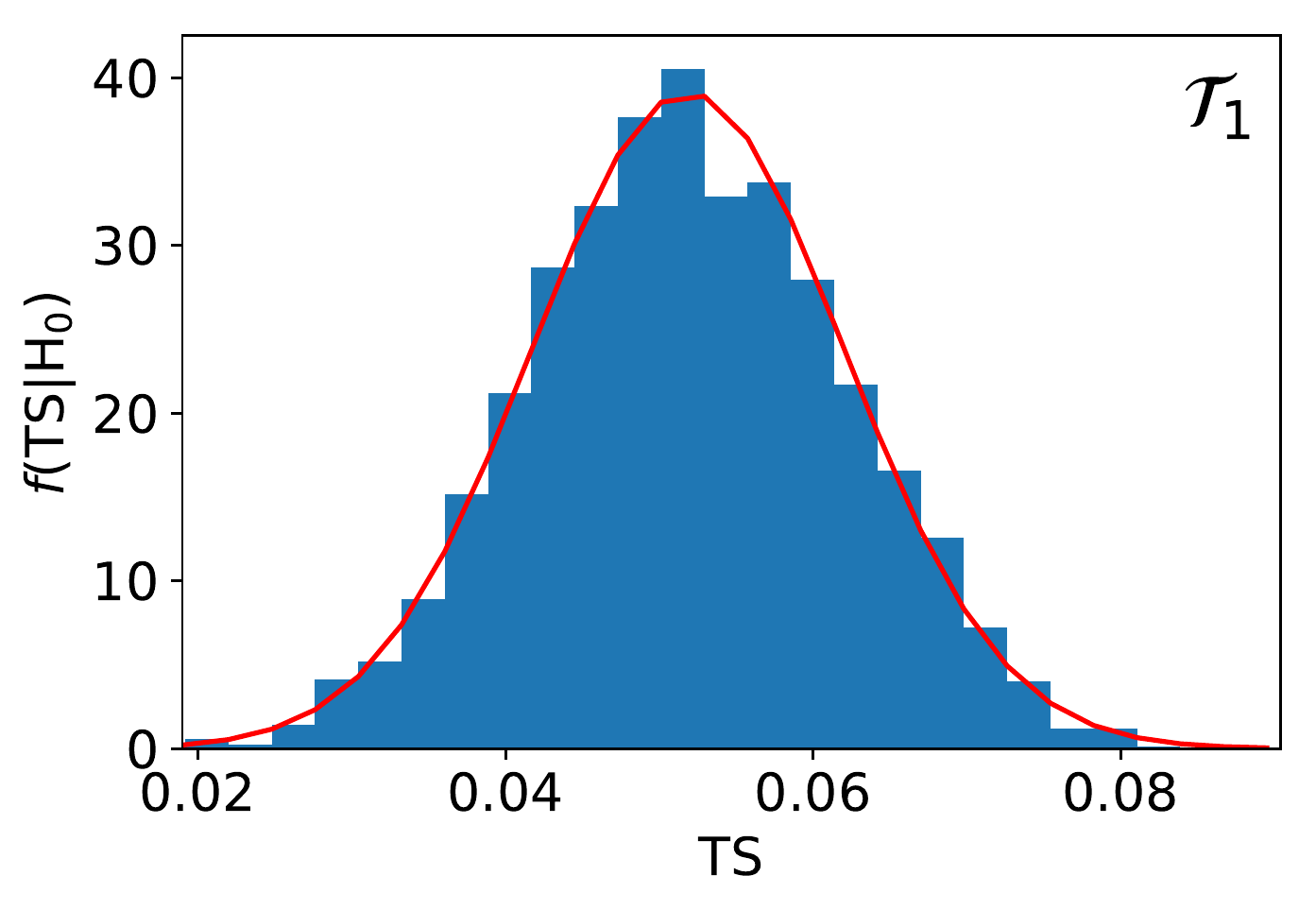}
\includegraphics[width=0.32\linewidth]{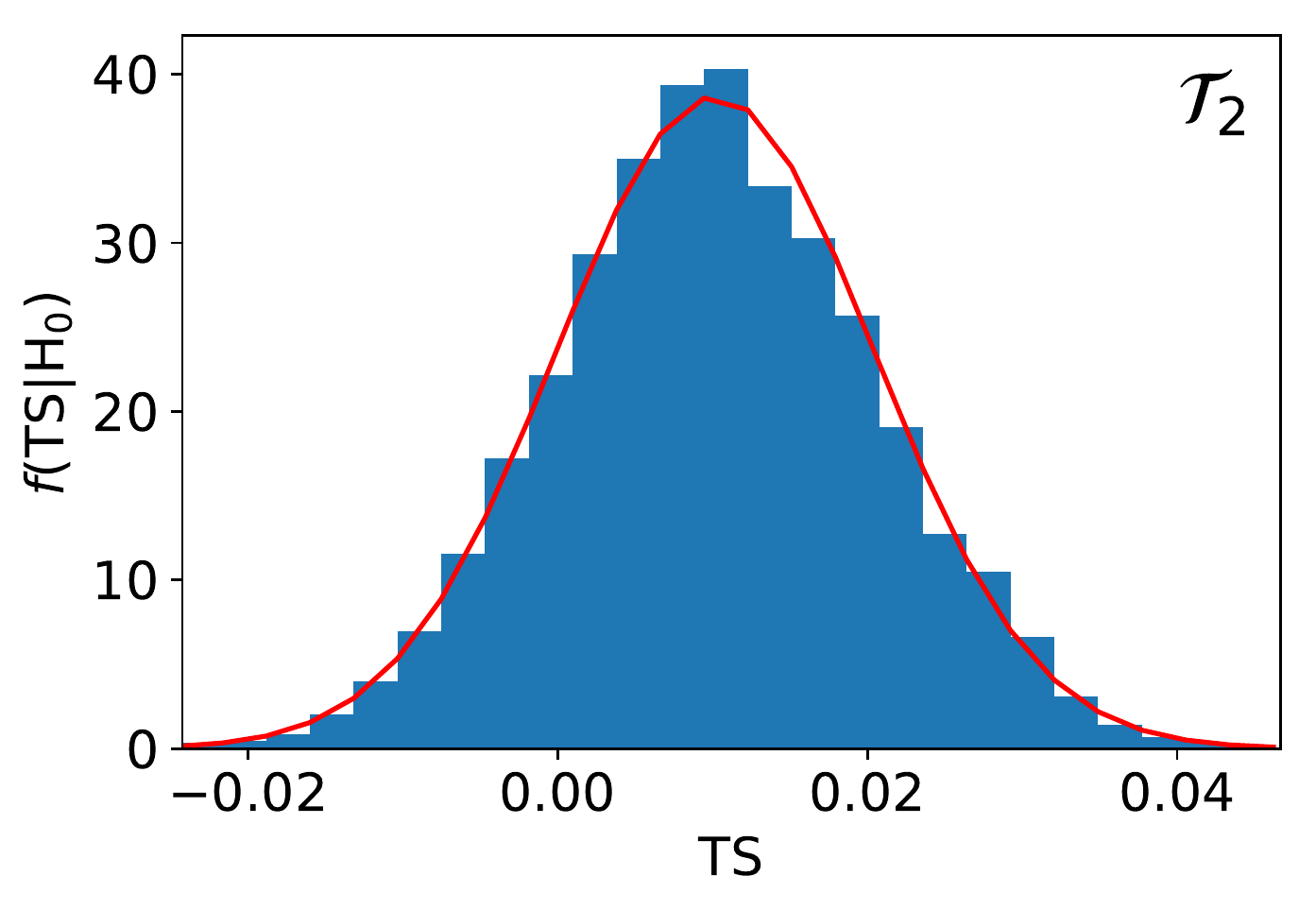}
\includegraphics[width=0.32\linewidth]{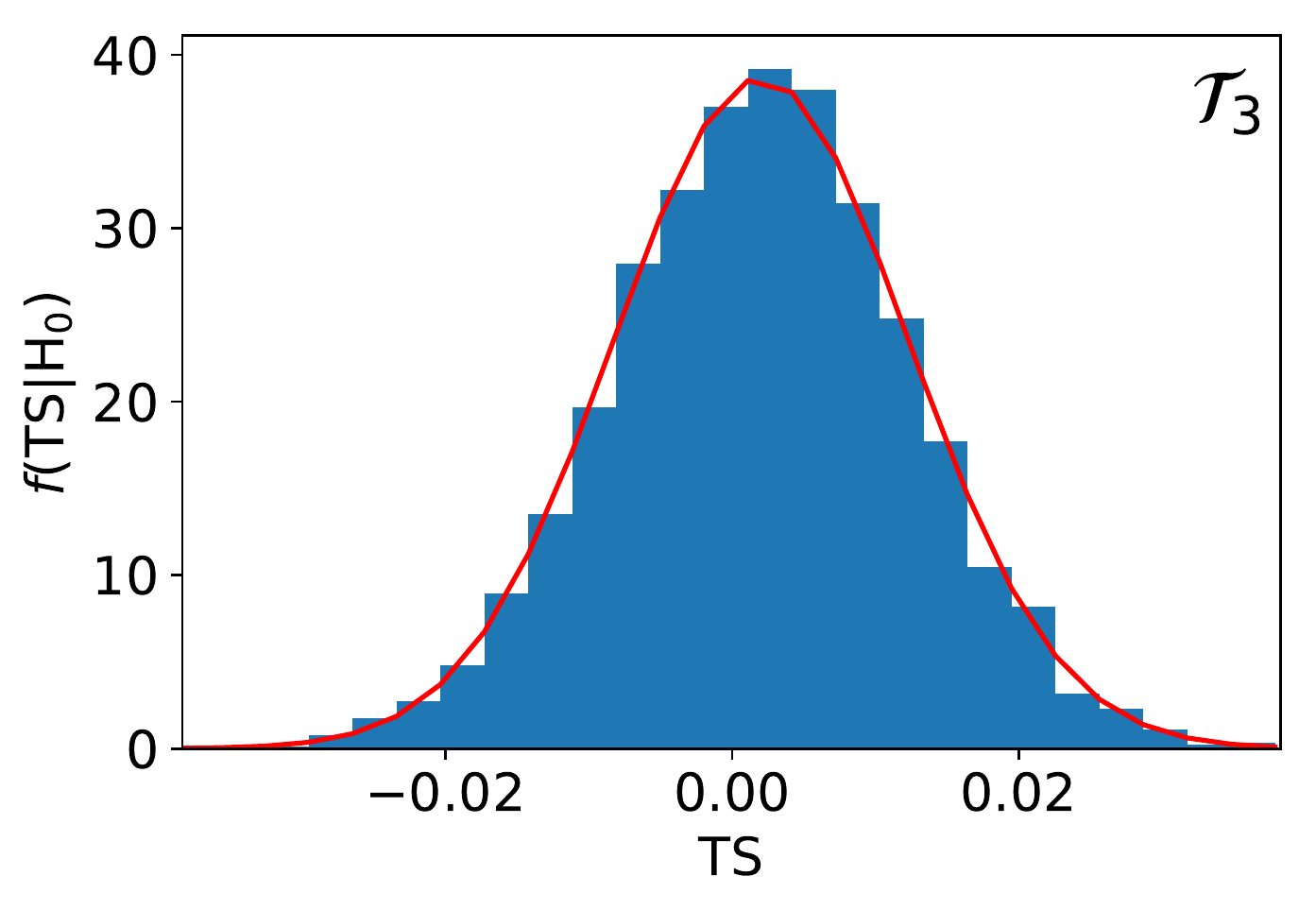}
\caption{
Distribution of the test statistic under the null hypothesis for our 3 signal points. Overlayed is a Gaussian distribution with the same mean and standard deviation as the data.
}
\label{fig:TS-monojet}
\end{figure}

The calculation of $\ts_{\rm obs}$ is performed for 100 random divisions.
The $p$-value and significance $Z$ of each $\ts_{\rm obs}$ are calculated with respect to the empirical distribution $f(\ts|H_0)$ where possible. In many cases, $\ts_{\rm obs}$ is so extreme that it falls outside the measured range of $f(\ts|H_0)$, in which case $p$ and $Z$ are determined from a Gaussian distribution with mean $\hat \mu$ and
variance $\hat\sigma^2$. This is equivalent to assuming that $f(\ts|H_0)$ is well-approximated by a Gaussian, which is true to a good approximation, as seen in Fig.~\ref{fig:TS-monojet}.
To be conservative, the technique is only considered sensitive to the signal if all simulated observations of TS exclude the null hypothesis, i.e. we show the minimum $Z$ significance (and maximum $p$-value). These results are shown in Table~\ref{tab:monojet-results}, where we see that the background-only hypothesis is strongly excluded for $\Tsample_1$ and $\Tsample_2$, even though these points are not yet excluded by traditional LHC searches. Bear in mind that this is a proof-of-concept, and real-world results are unlikely to be as clean, as discussed in Section~\ref{subsec:realdata}.

\begin{table}[t]
\centering
\begin{tabular}{|c|c|c|c|c|c|c|}
\hline
Sample & $M_{\rm med}$ & $\sigma_\Tsample$ [pb] & $\sigma_{\rm signal}$ [pb] &  max($p$-value) & min($Z$)  \\ 
\hline
$\Tsample_1$ &1.2 TeV& 223.0 & 20.4 &  $<10^{-50}$ & $>15\,\sigma$ \\
\hline
$\Tsample_2$ &2 TeV& 206.4 & 3.8 & $5.7\times 10^{-25}$ &  $ 10 \, \sigma$ \\
\hline
$\Tsample_3$ &3 TeV& 203.2 & 0.6 & 0.90 & $0.13\, \sigma$ \\
\hline
\end{tabular}
\caption{Summary of monojet results comparing $\Bsample$ (background only) with $\Tsample$ (background plus  DM signal). The cross section corresponding to the trial sample is simply given by $\sigma_{\Tsample}=\sigma_{\rm background}+\sigma_{\rm signal}$. The $p$-value and $Z$ statistic show the compatibility between $\Bsample$ and $\Tsample$; Large $Z$ indicates that $\Tsample$ is not consistent with the background-only hypothesis. Note that these results will be weakened by application of uncertainties (see text for details).}
\label{tab:monojet-results}
\end{table}

\subsubsection*{Inclusion of uncertainties}
To test the sensitivity of this technique to uncertainties and errors in the background simulation, we use the method outlined in Section~\ref{subsec:ext_uncert} to estimate the drop in significance when uncertainties are taken into account. Uncorrelated Gaussian noise with $\epsilon = 10\%$ (as defined in Section~\ref{subsec:2dgaussians}) is added to $\Bsample$, allowing the construction of $f(\ts_u|H_0)$ using $N_{\rm iter} = 1000$.
Note that while the primary result without uncertainties is agnostic as to how the overall background sample is divided between $\Bsample$ and $\Tsample$, this is not the case when applying uncertainties. We construct $f(\ts_u|H_0)$ by repeatedly applying different noise to the same $\Bsample$, and so $\Bsample$ and $\Tsample$ must be defined from the outset, leaving just one measurement of $\ts_{\rm obs}$, for a random draw of $\Bsample$ and the background component of $\Tsample$ from the overall pool of background simulations.
For $(\Tsample_1, \Tsample_2, \Tsample_3)$, we find that without noise $Z = (40, 13, 2.7)$. Note that as expected, these are larger than the minimum values over 100 observations reported in Table~\ref{tab:monojet-results}. With $\epsilon = 10\%$, we find that this reduces to $Z = (26, 12, 2.5)$ for the 3 samples, respectively. 
This is in line with expectations:  while this is a powerful technique, limited knowledge of the expected background will degrade the results. 
With this in mind, we reiterate that results based on simulations alone should be taken with a grain of salt. They show the strengths of the statistical test we are proposing and prove it is worthwhile to investigate it further, but they will be weakened 
in a real-world situation. 

As an application to experimental data, our technique could be applied by seeding the simulated background $\Bsample$ with noise associated with uncertainties in the Monte-Carlo background estimation, 
or seeding the measured data sample $\Tsample$ with noise
associated with systematic uncertainties.

\subsubsection*{Discussion}

\begin{figure}
\centering
\includegraphics[width=0.45\linewidth]{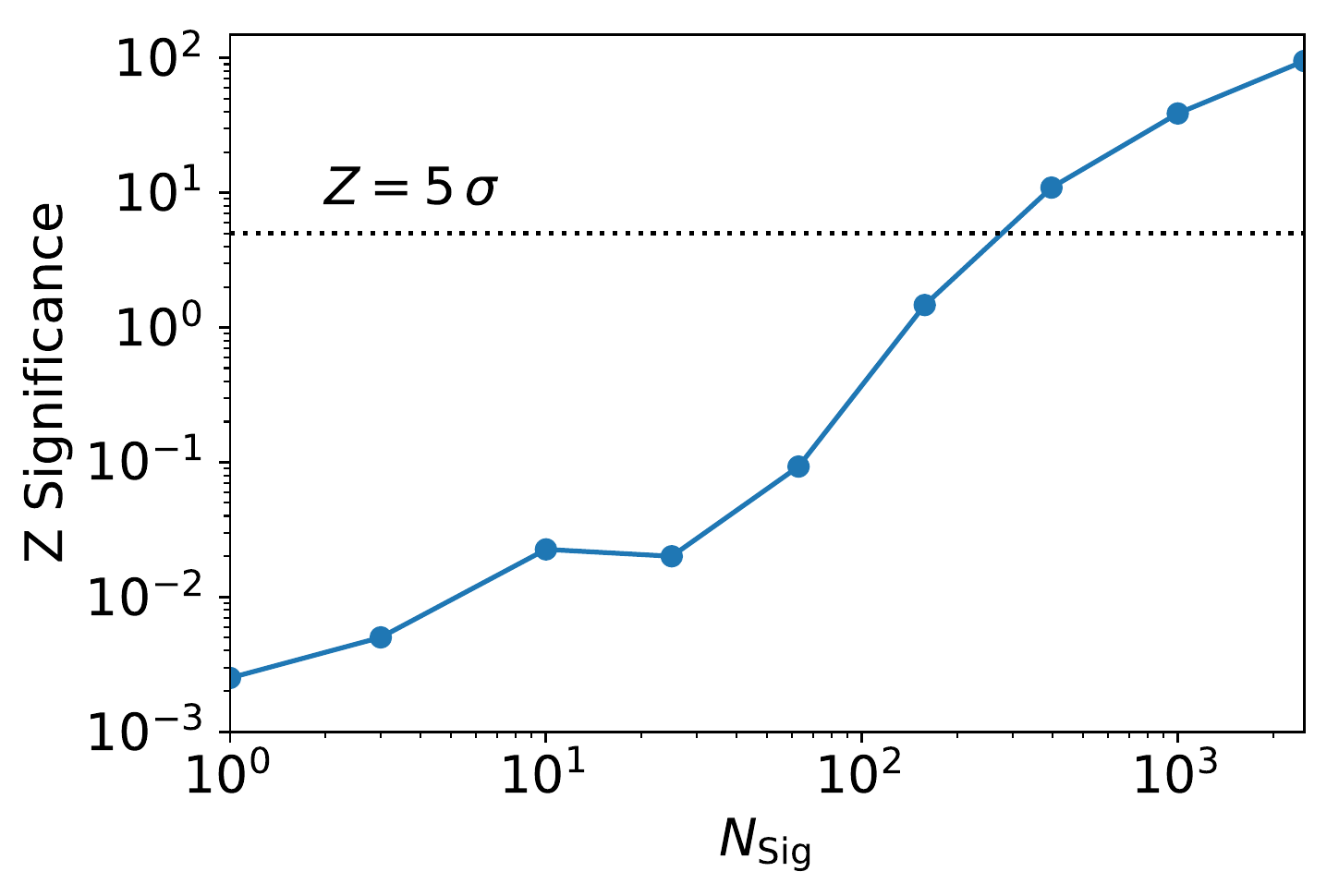}
\includegraphics[width=0.45\linewidth]{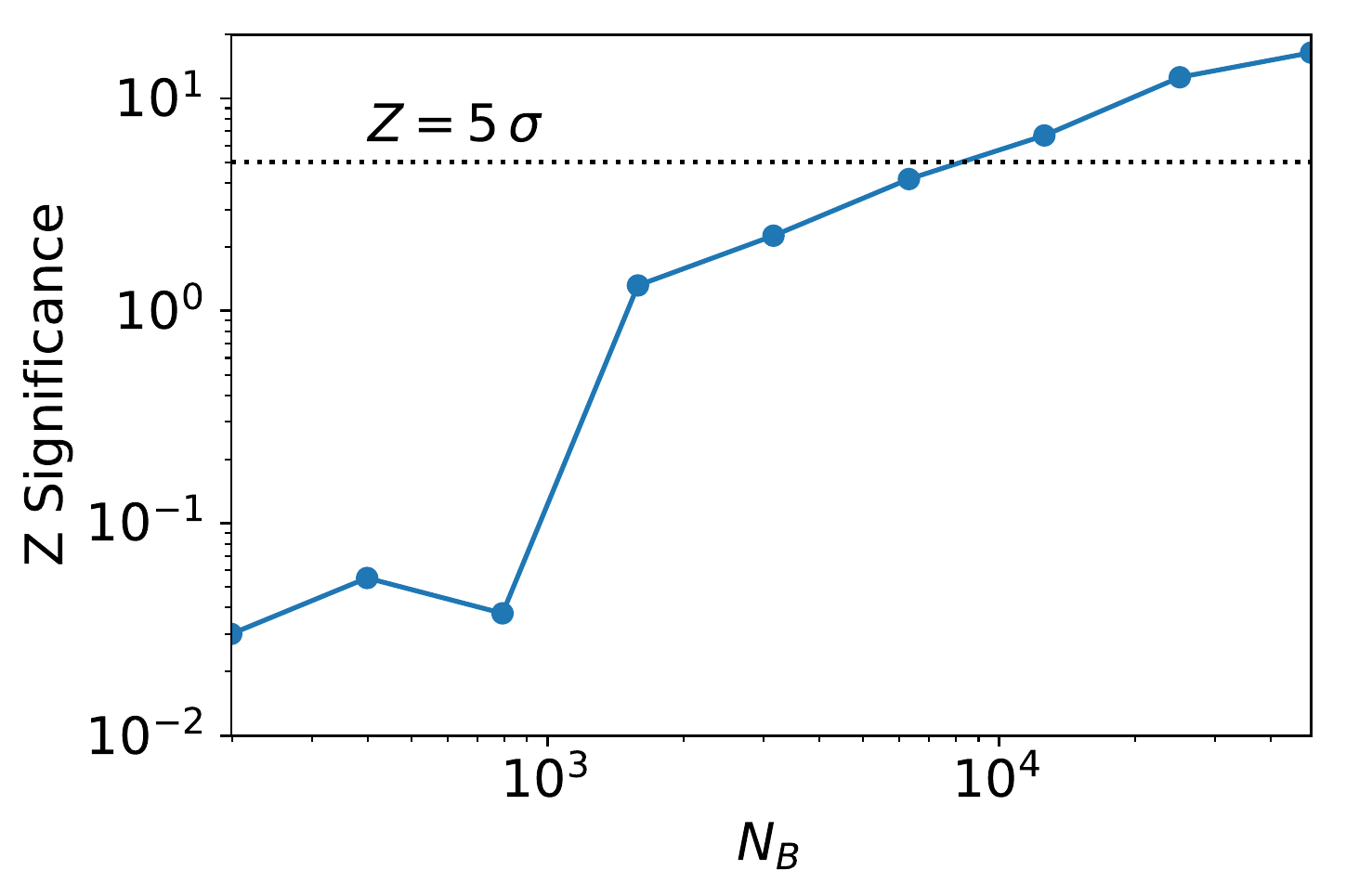}
\caption{The effect of $N_{\rm sig}$ (left) and $N_B$ (right) on the ability of the algorithm to distinguish $\Bsample$ and $\Tsample$. For the left figure, $N_B = 20\,000$ background events and $N_T = N_B + N_{\rm sig}$. (Based on the actual simulated signal and background cross-sections, the true value is $N_{\rm sig} = 375$.) In the right figure, $N_T = N_B + N_{\rm sig}$, where $N_{\rm sig}$ varies in proportion to $N_B$ and the relative signal/background cross-sections. In both cases, we use the trial sample $\Tsample_2$ corresponding to the signal with $M_{\rm med} = 2$ TeV.}
\label{fig:zvsX-monojet}
\end{figure}

To study the threshold to which this technique is sensitive, we can construct $\Tsample$ by adding an arbitrary number of signal events to the background, without reference to the relative signal cross-section. The result is shown in Figure \ref{fig:zvsX-monojet} (left panel), using the signal dataset with 
$M_{\rm med} = 2$ TeV. 
For each value of $N_{\rm sig}$, the distribution $f(\ts|H_0)$ is constructed over 1000 permutations, and the $Z$ significance is determined through taking the minimum value of $Z$ over 100 measurements of $\ts_{\rm obs}$ for different background permutations. There is a clear threshold, below which the significance is negligible and constant, and above which the significance grows as a power-law. The number of signal events in $\Tsample_2$ crosses this threshold while $\Tsample_3$ does not, explaining the rapid drop in the significance.

The strength of the technique is also sensitive to the number of samples. Figure~\ref{fig:zvsX-monojet} (right panel) demonstrates this, again using the signal dataset with $M_{\rm med} = 2$ TeV, $N_{\rm perm} = 1000$, and taking the minimum $Z$ over 100 measurements of $T_{\rm obs}$. It shows an approximately power-law growth in the significance, consistent with the same growth in the significance with number of signal events. Clearly, the more data the better.

\subsection{Future application to real data}
\label{subsec:realdata}

In a practical application of this technique by experimental collaborations, $\Bsample$ would correspond to simulations of the SM background, while $\Tsample$ would be the real-world observation, consisting of an unknown mix of signal and background events. Both $\Bsample$ and $\Tsample$ could be constructed under the same set of minimal cuts, imposed based on trigger requirements rather than as a guide to finding new physics.
While the technique itself is model-independent, there is freedom to apply physical knowledge in the choice of minimal cuts to keep the background simulation and data load manageable, and in the choice of feature vector, which can either be low-level (raw 4-vectors of reconstructed objects, or even pixel hits) or high-level (missing energy, hadronic energy etc.).

Even though we have only applied our method to a generic monojet signal, the strength of the algorithm is that it is sensitive to unspecified signals, and is limited only by the accuracy of the background simulation.
We emphasize that our case study in Section~\ref{subsec:monojet} is a proof of concept with a generic signal and a na\"ive estimation of the background. 

Accurately estimating SM backgrounds at the LHC is a significant challenge in the field and must be considered carefully in any future application of this technique. Currently used techniques of matching simulations to data in control regions still allow the use of our method, although this introduces some model-dependent assumptions. 
Alternatively, one may apply our statistical test in the context of data-driven background calculation, as a validation tool to measure the compatibility of Monte-Carlo simulations with data in control regions.

For instance, it is common practice to tune the nuisance parameters 
in order to make the Monte-Carlo simulation of the background match
the data in control regions. When one deals with more than one control region, this procedure results in a collection of patches of the feature space, 
in each of which the background simulation is fit to the data.
The statistical test we propose in this paper can be used to determine
to what extent (significance) the background simulation is representative of the data at the global level, in all control regions. 
And in case of discrepancies, it can pinpoint the regions of feature space where the mismatch between data and simulations is the largest.

As we have shown by implementing sample uncertainties in our statistical test,  
the test alone may not be sufficient to claim discovery in cases where background simulations are not sufficiently accurate, but this does not weaken the value of the method. It remains valuable as a tool to identify regions of excess in a model-independent way, allowing follow-up hand-crafted analyses of potential signal regions.

\section{Directions for extensions}
\label{sec:extensions}

In this section we summarize two main directions to extend and improve the method proposed in this paper.
We limit ourselves to just outlining some ideas, leaving a more complete analysis of each of these issues to future work.

\subsection{Adaptive choice of the number 
of nearest neighbors}
\label{subsec:adaptive}

The procedure for the density ratio estimator described in Section \ref{subsec:estimator} relies on choosing the number $K$ of NN.
As mentioned earlier, it is also possible to make the algorithm completely
unsupervised by letting it choose the optimal value of $K$.

One approach is to proceed by model selection as
in Refs.~\cite{SugiyamaMuller, SUGIYAMA2011735,kremer}.
We define the loss function as a mean-squared error between the true (unknown) density ratio $r(\boldsymbol{x})=p_T(\boldsymbol{x})/p_B(\boldsymbol{x})$ and the estimated density ratio
$\hat r(\boldsymbol{x})=\hat p_T(\boldsymbol{x})/\hat p_B(\boldsymbol{x})$
over the benchmark PDF $p_B(\boldsymbol{x})$, 
\bea
L(r,\hat r)&=&\frac{1}{2}\int \left[\hat r(\boldsymbol{x}')- r(\boldsymbol{x}')\right]^2 
p_B(\boldsymbol{x}') d\boldsymbol{x}'\\
&=&\frac{1}{2}\int \hat r(\boldsymbol{x}')^2 p_B(\boldsymbol{x}') d\boldsymbol{x}'
-\int \hat r(\boldsymbol{x}) p_T(\boldsymbol{x}) d\boldsymbol{x}
+\frac{1}{2}\int  r(\boldsymbol{x}')^2 p_B(\boldsymbol{x}') d\boldsymbol{x}'\,,
\eea
where the last term is constant and can be dropped, 
thus making the loss function independent of the unknown ratio $r(\boldsymbol{x})$.
The estimated loss function is obtained by replacing the expectations over the unknown PDF $p_B$  with the empirical averages
\be
\hat L(r,\hat r) = \frac{1}{2N_B}\sum_{\boldsymbol{x'}\in \Bsample}\hat r(\boldsymbol{x'})^2
-\frac{1}{N_T}\sum_{\boldsymbol{x}\in \Tsample}\hat r(\boldsymbol{x})\,.
\label{loss}
\ee
So, one can perform model selection by minimizing the estimated loss function \eqref{loss} with respect to the parameter $K$ and choosing this value of $K$ as the optimal one.
However, this procedure may be computationally intensive as it requires running the full algorithm several times (one for each different value of $K$).

Another approach is to implement the Point-Adaptive k-NN density estimator (PAk) \cite{Rodriguez1492,  Laio2018,  2018arXiv180210549D}, which is an algorithm to automatically find a compromise between  large variance  of the k-NN estimator (for small $K$), and large bias (for large $K$) due to variations of the density of points.

\subsection{Identifying the discrepant regions}
\label{subsec:location}

Suppose that after running the statistical test described in this paper one finds a $p$-value leading to a rejection of the null hypothesis, or at least for evidence of incompatibility between the original PDFs.
This means that the absolute value of the test statistic on the actual samples $|\ts_{\rm obs}|$ is large enough to deviate from zero significantly (to simplify the discussion, we assume in this subsection that $\ts_{\rm obs}>0$ and the distribution of $\ts$ has zero mean and unit variance: 
$\hat \mu=0, \hat\sigma =1$).
Then, our algorithm  has a straightforward by-product: it allows
to characterize the regions in feature space which contribute the most 
to a large $\ts_{\rm obs}$.

\begin{figure}[t]
\centering
\includegraphics[width=0.6\linewidth]{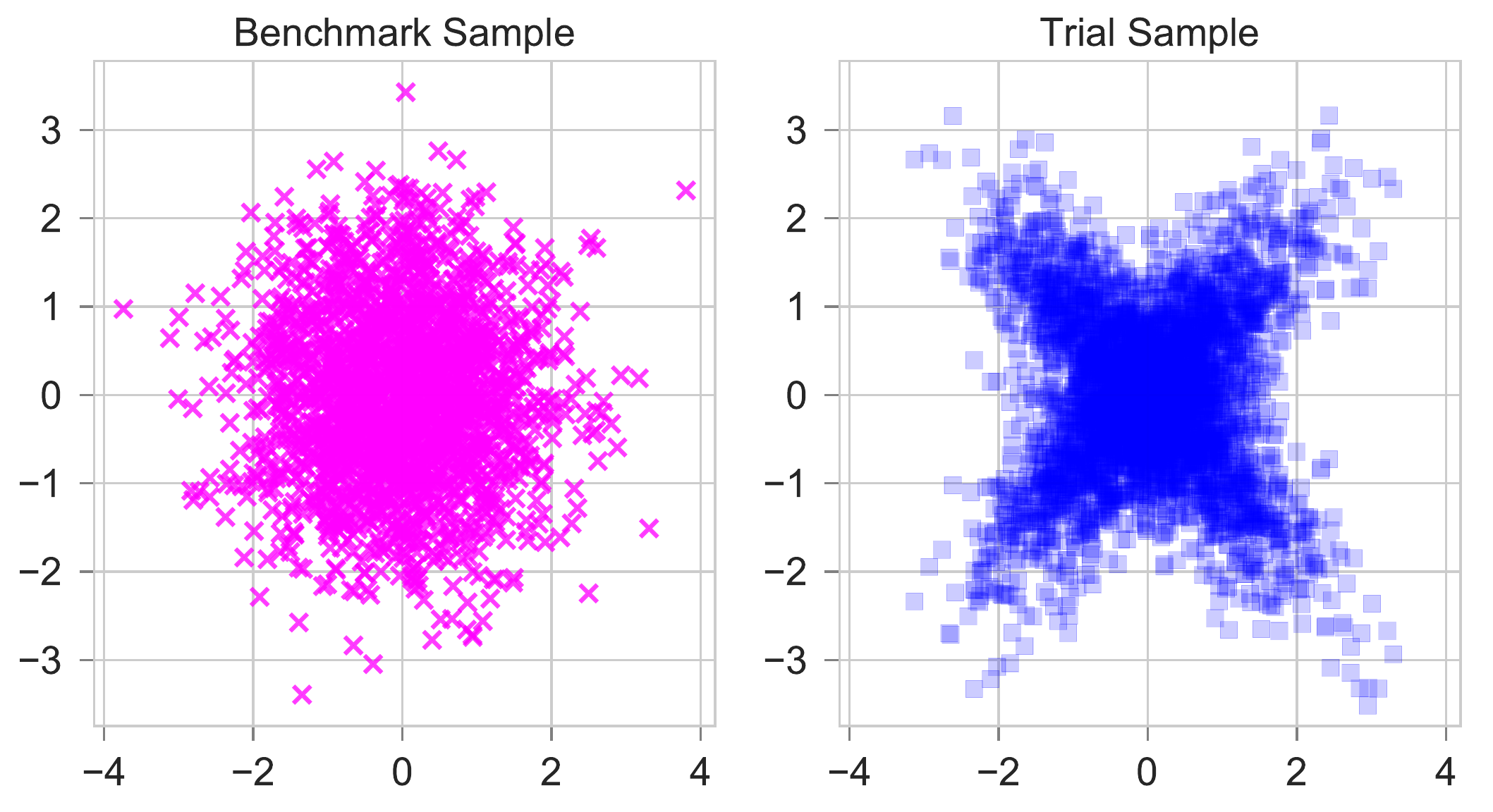}\\
\includegraphics[width=0.6\linewidth]{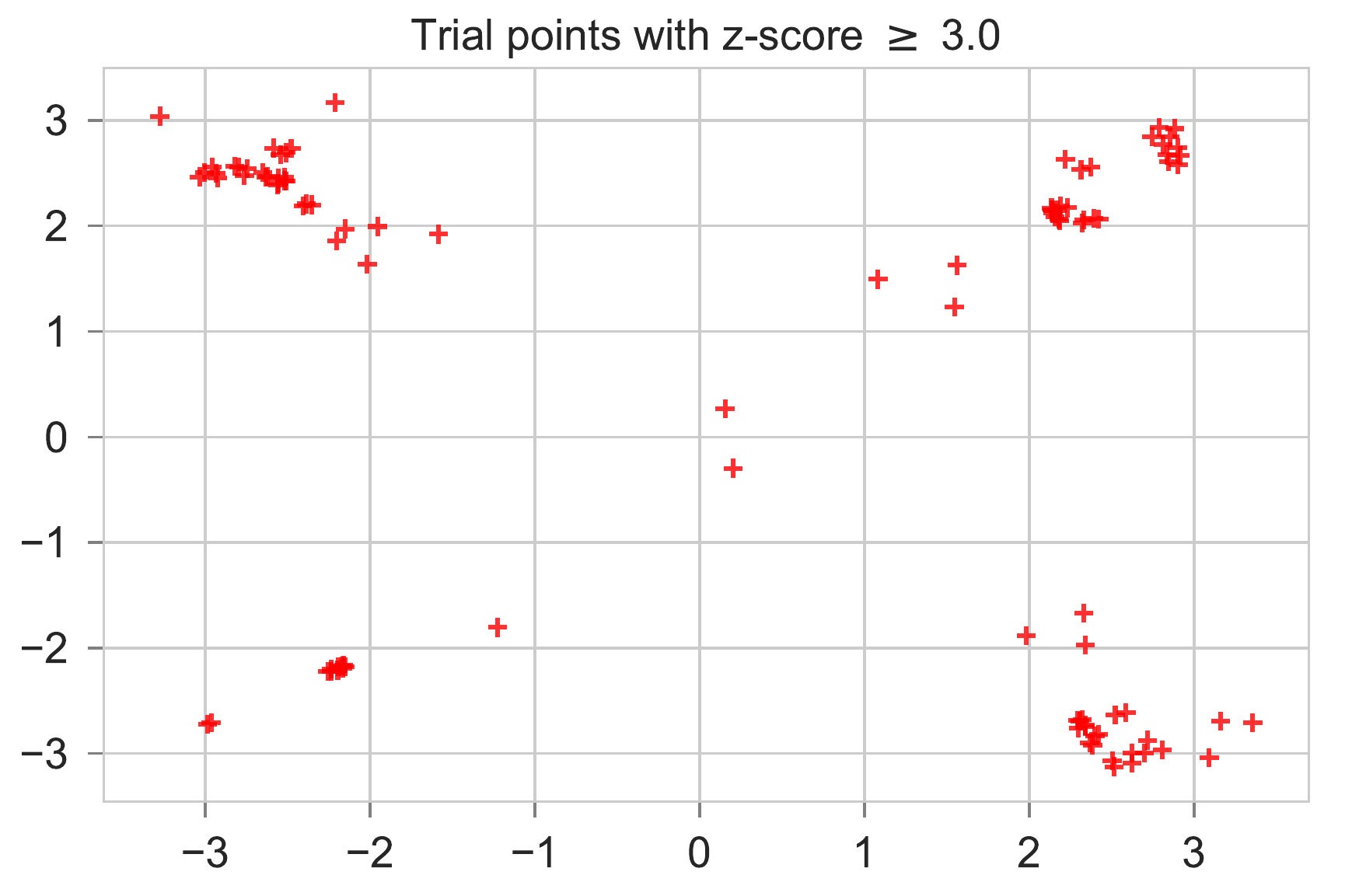}
\caption{
\textit{Upper panel:} benchmark (magenta crosses, left) and trial (blue squares, right) samples.
\textit{Lower panel:} points of trial sample with $z>3.0$;
this condition isolates the regions where most of the discrepancy between samples occurs.}
\label{fig:zfield}
\end{figure}

From the expression of the test statistic in Eq.~\eqref{eq:ts2} we see
that we may associate a density field $(\boldsymbol{x}_j)$ to each point 
$\boldsymbol{x}_j\in\Tsample$ as
\be
u(\boldsymbol{x}_j)\equiv \log\frac{r_{j,B}}{r_{j,T}}\,,
\ee
such that the test statistic is simply given by the expectation value 
(arithmetic average) of $u(\boldsymbol{x}_j)$ over the whole 
trial sample $\Tsample$
\be
\ts_{\rm obs} = D\cdot\textrm{E}_\Tsample[ u(\boldsymbol{x}_j)]
+ \log\frac{N_B}{N_T-1}\,.
\ee
It is then convenient to define a $z$-score field over the trial sample, 
by standard normalization of $u(\boldsymbol{x}_j)$ as
\be
z(\boldsymbol{x}_j)\equiv\frac{u(\boldsymbol{x}_j)
-\textrm{E}_\Tsample[ u(\boldsymbol{x}_j)]}{\sqrt{\textrm{Var}_\Tsample[u(\boldsymbol{x}_j)]}}\,.
\ee
One can then use this score field to identify those points in $\Tsample$
which are significantly larger than $\ts_{\rm obs}$, and they can be interpreted as the regions (or clusters) where the two samples manifest larger discrepancies.

This way, the $z$-score field provides a guidance for characterizing the regions in feature space where the discrepancy is more relevant, similar in spirit to regions of large signal-to-background ratio.
For instance, the points $\boldsymbol{x}_j$ with $z(\boldsymbol{x}_j)$ larger than a given threshold, e.g. $z(\boldsymbol{x}_j)>3$, are the points where one expects most of the ``anomaly'' to occur.
An example of this is shown in Figure \ref{fig:zfield}, 
where a circular $\Bsample$ sample is compared with a cross-like
$\Tsample$ sample. As expected, the $z$-field has higher density in  correspondence of the corners of the cross.

Such regions of highest incompatibility between trial 
and benchmark samples may even be clustered using standard clustering algorithms, thus extending the method studied in this paper with another unsupervised learning technique.

Once they have been characterized and isolated, these high-discrepancy regions in feature space can provide a guidance for further investigation, in order to identify what causes the deviations. For example, they can be used to place data selection cuts.

\section{Conclusions}
\label{sec:conclusion}

Many searches for new phenomena in physics (such as searches for New Physics at the LHC) rely on testing specific models and parameters.  Given the unknown nature of the physical phenomenon we are searching for, it is becoming increasingly important to find model-independent methods that are sensitive to an unknown signal hiding in the data. 

The presence of a new phenomenon in data manifests itself as deviations from the expected distribution of data points in absence of the phenomenon.
So, we propose a general statistical test for assessing the degree of compatibility between two datasets. 
Our method is model-independent and non-parametric, requiring no information about the parameters or signal spectrum of the new physics being tested; it is also un-binned, taking advantage of the full multi-dimensional feature space. 

The test statistic we employ to measure the `distance' between two datasets is built upon a 
nearest-neighbors estimation of their relative local densities. This is compared with the distribution of the test statistic under the null hypothesis. Observations of the test statistic at extreme tails of its distribution indicate that the two datasets come from different underlying probability densities.

Alongside an indication of the presence of anomalous events, our method can be applied to characterize the regions of discrepancy, providing a guidance for further analyses even in the case where one of the two samples (e.g. the background) is not known with enough accuracy to claim discovery.

The statistical test proposed in this paper has a wide range of scientific and engineering applications,  e.g.
to decide whether two datasets can be analyzed jointly, to find outliers in data, to detect changes of the underlying distributions over time, to detect anomalous events in time-series data, etc.

In particular, its relevance for particle physics searches at LHC is clear. In this case the observed data can be compared with simulations of the Standard Model in order to detect the presence of New Physics events in the data. Our method is highly sensitive even to a small number of these events, showing the strong potential of this technique.

\acknowledgments
We would like to thank A.~Davoli and A.~Morandini for collaboration at the early stages of this work, and  D.~Barducci, R.~Brasselet, R.T.~D' Agnolo, A.~Farbin, F.~Gieseke, E.~Merelli, E.~Mer\'enyi, A.~Laio, L.~Lista and A.~Wulzer for insightful discussions.

\appendix

\section{Kullback-Leibler divergence}
\label{app:KL}

The Kullback-Leibler (KL) divergence (or distance) is one of the most fundamental 
measures in information theory. The KL divergence of two continuous probability density functions (PDF) $P,Q$ is defined as
\be
D_{\rm KL}(P||Q)\equiv \int P(\boldsymbol{x})
\log\frac{P(\boldsymbol{x})}{Q(\boldsymbol{x})}
d\boldsymbol{x}\,,
\label{KLexact}
\ee
and it is a special case of $f$-divergences. 

If the distributions $P,Q$ are not known, but we are only given two samples $\mathcal{P}=\{\boldsymbol{x}_i\}_{i=1}^{N_P}$ of i.i.d.
points drawn from $P$ and
$\mathcal{Q}=\{\boldsymbol{x}_i'\}_{i=1}^{N_Q}$ of i.i.d. points
drawn from $Q$, it is possible to estimate the KL divergence using
empirical methods. 
The estimated KL divergence between the estimated PDFs of $\hat P, \hat Q$
is obtained by replacing the PDFs $P,Q$ with their estimates $\hat P,\hat Q$ and 
replacing the expectation value in Eq.~\eqref{KLexact} with the empirical (sample) average 
\be
\hat D_{\rm KL}(\hat P||\hat Q) =\frac{1}{N_P}\sum_{j=1}^{N_P}
\log\frac{\hat P(\boldsymbol{x}_j)}{\hat Q(\boldsymbol{x}_j)}\,.
\label{KLestimated}
\ee
For the special case of Gaussian PDFs, the calculation of the KL divergence is particularly simple. 
Given two multivariate ($D$-dimensional) Gaussian PDFs
defined by mean vectors $\boldsymbol{\mu}_{1,2}$
and covariance matrices $\Sigma_{1,2}$:
\be
P=\mathcal{N}(\boldsymbol{\mu}_1, \Sigma_2)\,,
\qquad
Q=\mathcal{N}(\boldsymbol{\mu}_2, \Sigma_2)\,,
\ee
the KL divergence in Eq.~\eqref{KLexact} is given by
\be
D_{\rm KL}(P||Q)=\frac{1}{2}\left[
(\boldsymbol{\mu}_2-\boldsymbol{\mu}_1)^T \Sigma_2^{-1}
(\boldsymbol{\mu}_2-\boldsymbol{\mu}_1)
+\textrm{Tr}(\Sigma_2^{-1} \Sigma_1)
+\log\frac{\det\Sigma_2}{\det\Sigma_1}-D
\right]\,.
\label{KL_gaussians}
\ee

\bibliographystyle{JHEP} \bibliography{bibliography} 
\end{document}